\documentstyle[12pt]{article}

\def\theequation{\arabic{section}.\arabic{equation}}

\def\thesubsection    {\Alph{subsection}}

\def\thefootnote{\fnsymbol{footnote}}

\def\be{\begin{equation}}
\def\ee{\end{equation}}
\def\bea{\begin{eqnarray}}
\def\nn{\nonumber}
\def\eea{\end{eqnarray}}

\def\taderiv{\hspace{0.7ex}
  \tilde{\mbox{ }}
  \hspace{-1.5ex}\bigtriangledown}

\def\tderiv{\hspace{0.4ex}
  \tilde{\mbox{ }}
  \hspace{-2ex}\bigtriangledown}

\def\tBox{\mbox{}^{\tilde{\hspace{1ex}}}
  \hspace{-1.4ex}\Box}

\begin{document}


\begin{titlepage}


\baselineskip 7ex
\mbox{}\vspace*{1.5ex}

\begin{center}
\Large {\bf An effective stochastic semiclassical theory for the
gravitational field}
\end{center}

\vspace{2ex}

\baselineskip 5ex

\begin{center}
Rosario Mart\'{\i}n and Enric Verdaguer\footnote{Institut de 
      F\'{\i}sica d'Altes Energies (IFAE)} \\
{\it Departament de F\'{\i}sica Fonamental, 
Universitat de Barcelona, 
Av.~Diagonal 647}, \\
{\it \mbox{08028 Barcelona}, Spain}
\end{center}

\vspace{7ex}

\begin{abstract}
Assuming that the mechanism proposed by Gell-Mann and Hartle works as
a mechanism for decoherence and classicalization of the metric field,
we formally derive the form of an effective theory for the
gravitational field in a semiclassical regime. This effective theory
takes the form of the usual semiclassical theory of gravity, based on
the semiclassical Einstein equation, plus a stochastic correction
which accounts for the back reaction of the lowest order matter 
stress-energy fluctuations. 
\end{abstract}
\vspace{6ex}

\end{titlepage}


\section{\hspace{-2.5ex}. Introduction}
\label{sec:introduction}


\setcounter{equation}{0}

In the semiclassical theory of gravity, the gravitational field is
treated classically, but the matter fields are quantum. The key
equation of the theory is the semiclassical Einstein equation,
a generalization of the Einstein equation where the expectation value 
of the stress-energy tensor of quantum matter fields is the source of
curvature. 

One expects that semiclassical gravity could be derived from a
fundamental quantum theory of gravity as a certain approximation,
but, in the absence of such a fundamental theory, the scope and
limits of the semiclassical theory are not very well understood. 
It seems clear, nevertheless, that it
should not be valid unless gravitational fluctuations  
are negligibly small. 
This condition may break down when the matter stress-energy has
appreciable quantum fluctuations,
since one would expect that fluctuations in the stress-energy of
matter would induce gravitational fluctuations \cite{ford82}. 
A number of examples have been recently studied, both in cosmological
and flat spacetimes, where, for some states 
of the matter fields, the stress-energy tensor have significant
fluctuations \cite{stress-en_fluctu}. 
To account for such fluctuations, it is necessary to extend the 
semiclassical theory of gravity. 

To address this problem, or analogous problems in quantum mechanics or
quantum field theory, different approaches have been adopted in the
literature. The present paper attempts to unify, at
least conceptually, two of these approaches in a formal derivation of
an effective theory for the gravitational field in the semiclassical
regime. The common feature of these two approaches is the idea of
viewing the metric field as the system of interest and the matter
fields as being part of its environment. This idea was first proposed
by Hu \cite{hu89} in the context of semiclassical cosmology.    
Both approaches make use of the influence functional formalism,
introduced by Feynman and Vernon \cite{feynman-vernon} to deal with
a system-environment interaction in a full quantum
theory. In this formalism, the integration of the environmental
variables in a path integral yields the influence functional, from
which one can define an effective action for the dynamics of the
system \cite{feynman-hibbs,calzettahu,humatacz,husinha,%
caldeira,hu-paz-zhang,hu-matacz94,greiner}. 

The first of these two approaches has been extensively used in the
literature, not only in the framework of semiclassical cosmology
\cite{calzettahu,humatacz,husinha,cv96,lomb-mazz,ccv97,campos-hu}, but
also in the context of analogous semiclassical regimes for systems of
quantum mechanics \cite{caldeira,hu-matacz94,hu-paz-zhang2}
and of quantum field theory 
\cite{greiner,matacz,morikawa,shaisultanov,gleiser}.
It makes use of the closed time path (CTP) functional technique, due
to Schwinger and Keldysh \cite{schwinger}.
This is a path integral technique designed to obtain expectation
values of field operators in a direct way \cite{ctp}.
In the semiclassical regime, a tree level approximation is performed
in the path integrals involving the system variables. In this
approximation, the equation of motion for the expectation value of the
system field operator is the semiclassical equation, which can be
directly derived from the effective action of Feynman and Vernon
\cite{calzettahu,greiner,cv96,ccv97,campos-hu,shaisultanov}.
When computing this effective action perturbatively up to quadratic
order in its variables, one usually finds some imaginary terms which
do not contribute to the semiclassical equation. The key point of this
approach is the formal identification of the contribution of such
terms to the influence functional with the characteristic functional of
a Gaussian stochastic source. Assuming that in the semiclassical
regime this stochastic source interacts with the system variables, 
and, thus, these become stochastic variables, equations of the
Langevin type are derived for these variables. 
However, since this approach relies on a purely formal identification,
doubts can be raised on the physical meaning of the derived equations.

The second approach is based on the description of the transition from
quantum to classical behavior in the framework of the consistent
histories formulation of a quantum theory. 
The consistent histories formulation, proposed by Griffiths
\cite{griffiths}, and developed by Omn\`es \cite{omnes} and by
Gell-Mann and Hartle \cite{gell-mann-hartle,hartle}, 
was designed to deal
with quantum closed ({\it i.e.}, isolated) systems. It is thus
believed to be an appropriate approach to quantum cosmology, where the
quantum system is the whole universe. The main goal of this
formulation is the study of the conditions under which a set of
quantum mechanical variables become decoherent, which means that 
these variables can be described in a probabilistic way
\cite{gell-mann-hartle,hartle,halliwell93,histories,paz-zurek}.
When the closed system consists on a distinguished subsystem 
(the ``system'', which is also often called an ``open system'')
interacting with its environment, Gell-Mann and Hartle proposed a
mechanism for decoherence and classicalization of suitably
coarse-grained system variables \cite{gell-mann-hartle,hartle}.
This approach allows to evaluate the probability distribution
functional associated to such decoherent variables and, under some
approximations, to derive effective quasiclassical equations of motion
of the Langevin type for such variables 
\cite{gell-mann-hartle,hartle,halliwell93,dowker,halliwell}.

In Sec.~\ref{sec:classicalization} we show that that these two approaches
can in fact be related. 
In this way, we see that, on the one hand, the second approach sheds
light into the physical meaning of the first one. On the other hand,
the first approach provides a tool for computing effective
Langevin-type equations to the second one.
A large portion of this section consists of reformulating 
the mechanism for decoherence and classicalization
of Gell-Mann and Hartle in the language of the CTP functional
formalism.  
  
In Sec.~\ref{sec:Einstein-Langevin}, we use the results of this
analysis to formally derive effective equations of motion for the
gravitational field in a semiclassical regime. 
This derivation relies heavily on the results of the previous section.
We find that, in the semiclassical 
regime, gravity might be described by a background metric, solution of
the semiclassical Einstein equation, plus some stochastic metric
perturbations. The equation for these perturbations, the semiclassical
Einstein-Langevin equation, is seen to incorporate the effect of the
lowest order matter stress-energy fluctuations on the gravitational
field.  

In this paper we use the $(+++)$ sign conventions and the abstract index
notation of Ref.~\cite{wald84}, and we work in units in which
$c=\hbar =1$.

\newpage


\section{\hspace{-2.5ex}. Effective equations of motion from 
environment-in\-duced classicalization}
\label{sec:classicalization}


\setcounter{equation}{0}
\setcounter{footnote}{0}

\def\thefootnote{\arabic{footnote}}


\subsection{\hspace{-2.5ex}. The CTP functional formalism 
for a system-environment interaction}
\label{subsec:CTP}


We start this section by sketching the CTP functional formalism
\cite{schwinger}
applied to a system-environment interaction and its relation with the
influence functional formalism of Feynman and Vernon
\cite{feynman-vernon}. For more
detailed reviews of the CTP functional formalism, see
Refs.~\cite{ctp,campos-hu}, 
and for the influence functional formalism of Feynman
and Vernon, see 
Refs.~\cite{feynman-hibbs,calzettahu,humatacz,husinha,%
caldeira,hu-paz-zhang,hu-matacz94,greiner}. 
For simplicity, we shall work in this section with a model of quantum
mechanics, but all the formalism can also be formally applied to field
theory. It is instructive to maintain in this section the explicit
dependence on $\hbar$. Let us consider a model of quantum mechanics
which describe the interaction 
of two subsystems: one, called the ``system'', with coordinates $q$,
and the other, called the ``environment'', with coordinates 
$Q$.\footnote{Even if, in order to simplify the notation, we do not write
indices in these coordinates, $q$ and $Q$ have to be understood as
representing an arbitrary number of degrees of freedom (which, in
particular, can be an infinite number of degrees of freedom).}
We write the action for this model as 
$S[q,Q]=S_s[q]+S_{se}[q,Q]$.\footnote{We shall assume 
that the action $S[q,Q]$ is the one that
appears in the path integral formulas for the model, which, in
general, needs not to coincide with the classical action for the model
\cite{abers,weinberg}.}
Let $\hat{q}(t)$ and $\hat{Q}(t)$ be the Heisenberg picture coordinate
operators, which are assumed to be self-adjoint, 
{\it i.e.}, $\hat{q}^{\dag}\!=\!\hat{q}$ and 
$\hat{Q}^{\dag}\!=\!\hat{Q}$, and
let $\hat{q}^{\rm \scriptscriptstyle S}$ and 
$\hat{Q}^{\rm \scriptscriptstyle S}$ be the corresponding 
Schr\"{o}dinger picture operators.
Suppose that we are only interested in describing the 
physical properties of system observables from
some initial time $t_i$ until some final time $t_f>t_i$.
Working in the
Schr\"{o}dinger picture, the state of the full system 
({\it i.e.}, system plus environment) at the initial time
$t\!=\!t_i$ will be described by a density operator
$\hat{\rho}^{\rm \scriptscriptstyle S}(t_i)$. 
Let $\left\{ |q,Q\rangle^{\rm \scriptscriptstyle S} \right\}$ 
be the basis of
eigenstates of the operators $\hat{q}^{\rm \scriptscriptstyle S}$ 
and $\hat{Q}^{\rm \scriptscriptstyle S}$. The
matrix elements of the initial density operator in this basis will be
written as  
$\rho(q,Q;q^{\prime},Q^{\prime};t_i)\equiv 
\mbox{}^{\rm \scriptscriptstyle S}
\langle q,Q|\:
\hat{\rho}^{\rm \scriptscriptstyle S}(t_i)
\:|q^{\prime},Q^{\prime}\rangle^{\rm \scriptscriptstyle S}$. 
For simplicity, we shall assume that the initial density operator can
be factorized as 
$\hat{\rho}^{\rm \scriptscriptstyle S}(t_i) \!=\!
\hat{\rho}_s^{\rm \scriptscriptstyle S}(t_i)\otimes 
\hat{\rho}_e^{\rm \scriptscriptstyle S}(t_i)$,
in such a way that its matrix elements in coordinate representation
can be written as 
$\rho(q,Q;q^{\prime},Q^{\prime};t_i)\!=\!\rho_s(q,q^{\prime};t_i)\,
\rho_e(Q,Q^{\prime};t_i)$. 
However, the formalism can be generalized to the
most general case of a non-factorizable initial density operator
\cite{hakim,grabert,gell-mann-hartle}. 
We are interested in computing 
expectation values of operators related to
the system variables only, for times $t$ between $t_i$ and $t_f$.
The dynamics of the system in this sense can be completely
characterized by the knowledge of the whole family of Green functions
of the system. Working in the Heisenberg picture, these Green functions
can be defined as expectation values of products of $\hat{q}(t)$
operators. These Green functions can be derived from a CTP generating
functional in which only the system variables are coupled to external
sources $j_+(t)$ and $j_-(t)$ 
\cite{calzettahu,greiner,cv96,campos-hu,morikawa,shaisultanov}. 
This CTP generating functional can be written as the following
path integral\footnote{A way of generalizing 
the formalism to a non-factorizable
initial density operator consists in the following 
\cite{hakim,gell-mann-hartle}. 
One writes the initial density matrix in coordinate representation as 
$\rho(q,Q;q^{\prime},Q^{\prime};t_i)=\rho_s(q,q^{\prime};t_i)\,
\rho_{se}(q,Q;q^{\prime},Q^{\prime};t_i)$, where $\rho_s$ is chosen in
such a way that
$\int\! dq\, \rho_s(q,q;t_i)=1$. Then, the CTP generating functional
can be written as (\ref{generating functional}), with
\[
e^{{i \over \hbar}\,S_{\rm eff}[q_+,q_-]}\equiv 
\int\! {\cal D}[Q_+]\;{\cal D}[Q_-]\;
\rho_{se} (q_{+_{\scriptstyle i}},Q_{+_{\scriptstyle i}};
         q_{-_{\scriptstyle i}},Q_{-_{\scriptstyle i}};t_i ) \:
\delta(Q_{+_{\scriptstyle f}}\!-Q_{-_{\scriptstyle f}}) \;
e^{{i \over \hbar}\, \left(\,S[q_+,Q_+]-S[q_-,Q_-]\, \right)}.
\]
}
\be
Z[j_+,j_-] = \int\! {\cal D}[q_+]\;{\cal D}[q_-]\;
\rho_s (q_{+_{\scriptstyle i}},q_{-_{\scriptstyle i}};t_i ) \:
\delta(q_{+_{\scriptstyle f}}\!-q_{-_{\scriptstyle f}}) \;
e^{{i \over \hbar}\, \left(S_{\rm eff}[q_+,q_-]+ 
 \hbar \!\int\! dt\, j_+ q_+ -\hbar \!\int\! dt\, j_- q_- \right)},
\label{generating functional}
\ee
with
\be
S_{\rm eff}[q_+,q_-]\equiv S_s[q_+]-S_s[q_-]+S_{\rm IF}[q_+,q_-],
\label{effective action}
\ee
where $S_{\rm IF}$ is the influence action of Feynman and Vernon,
which is defined in terms of the influence functional 
${\cal F}_{\rm IF}$ as
\bea
\hspace*{-4.5ex}
{\cal F}_{\rm IF}[q_+,q_-]\!&\!\!\equiv \!\!&\! e^{{i \over \hbar}\,
  S_{\rm IF}[q_+,q_-]}  \nn \\
\!&\!\!\equiv \!\!& \!
\int\! {\cal D}[Q_+]\;{\cal D}[Q_-]\;
\rho_e (Q_{+_{\scriptstyle i}},Q_{-_{\scriptstyle i}};t_i ) \:
\delta(Q_{+_{\scriptstyle f}}\!-Q_{-_{\scriptstyle f}}) \;
e^{{i \over \hbar}\, \left(\,S_{se}[q_+,Q_+]-S_{se}[q_-,Q_-]\, \right)}.
\label{influence functional}
\eea
We shall call $S_{\rm eff}[q_+,q_-]$ the effective action of Feynman
and Vernon. 
In these expressions we use the notation 
$q_{+_{\scriptstyle i}} \!\!\equiv\! q_+(t_i)$, 
$q_{+_{\scriptstyle f}} \!\!\equiv\! q_+(t_f)$,
$Q_{+_{\scriptstyle i}} \!\!\equiv\! Q_+(t_i)$, 
$Q_{+_{\scriptstyle f}} \!\!\equiv\! Q_+(t_f)$, and similarly for $q_-$
and $Q_-$. All the integrals in $t$, including those that would define
the actions $S_s[q]$ and $S_{se}[q,Q]$ in terms of the corresponding
Lagrangians, have to be understood as integrals between $t_i$ and
$t_f$. The CTP generating functional has the properties
\be
Z[j,j]=1, \hspace{7 ex} 
Z[j_-,j_+]=Z^{\displaystyle \ast}[j_+,j_-], \hspace{7 ex}
\bigl|\hspace{0.2ex} Z[j_+,j_-]\hspace{0.2ex}\bigr|\leq 1.
\label{generating funct properties}
\ee
From this generating functional, we can derive the following Green
functions for the system:
\be
\left\langle\, \tilde{\rm T}[\hat{q}(t_1^{\prime}) \cdots
\hat{q}(t_s^{\prime})] \,{\rm T}[\hat{q}(t_1) \cdots
\hat{q}(t_r)]\, \right\rangle \hspace{-0.2ex}=\hspace{-0.2ex}
\left. {\delta Z[j_+,j_-]  \over 
i\delta j_+(t_1) \cdots i\delta j_+(t_r) 
(-i)\delta j_-(t_1^{\prime}) \cdots
(-i)\delta j_-(t_s^{\prime}) }\right|_{j_\pm=0} \!,
\label{green functions}
\ee
where $t_1,\dots ,t_r, t_1^{\prime},\dots ,t_s^{\prime}$ are all
between $t_i$ and $t_f$,  
${\rm T}$ and $\tilde{\rm T}$ mean, respectively, time and
anti-time ordering. The expectation value is taken in the
Heisenberg picture state corresponding to the Schr\"{o}dinger picture
state described by $\hat{\rho}^{\rm \scriptscriptstyle S}(t_i)$ at the
initial time $t\!=\!t_i$. The influence functional
(\ref{influence functional}) can actually be interpreted as a 
CTP generating functional for quantum variables $Q$ coupled to 
classical time-dependent sources $q(t)$ through the action
$S_{se}[q,Q]$ \cite{su}. 
Let us consider the quantum theory for the variables $Q$
in presence of classical sources $q(t)$ corresponding to this action,
and assume that the initial Schr\"{o}dinger picture state
for the quantum variables $Q$ is described by the density operator
$\hat{\rho}_e^{\rm \scriptscriptstyle S}(t_i)$. For this theory, 
let $\hat{\cal U}[q](t,t^{\prime })$ be the unitary
time-evolution operator, which can be formally written as  
$\hat{\cal U}[q](t,t^{\prime })\!=\! {\rm T} \exp \!
\left[-{i\over \hbar} \int_{t^{\prime }}^{t} dt^{\prime\prime }
\hat{H}^{\rm \scriptscriptstyle S}[q](t^{\prime\prime})\right]$,
for $t\!>\!t^{\prime }$, 
where $\hat{H}^{\rm \scriptscriptstyle S}[q](t)$ is
the Hamiltonian operator in the Schr\"{o}dinger picture. This
Hamiltonian operator depends on $t$ as a function of $q(t)$ and 
their derivatives 
$\dot{q}(t)$, and this gives a functional dependence on $q$
in the operator $\hat{\cal U}$. It is easy to see that 
\cite{gell-mann-hartle,hartle,humatacz,hu-matacz94,greiner,hakim}
\be
{\cal F}_{\rm IF}[q_+,q_-]=
{\rm Tr}  \Bigl[ \hat{\rho}_e ^{\rm \scriptscriptstyle S}(t_i)
\: \hat{\cal U}^{\dag}[q_-](t_{f},t_{i})\:
\hat{\cal U}[q_+](t_{f},t_{i})\Bigr]=
\Bigl\langle  \hat{\cal U}^{\dag}[q_-](t_{f},t_{i})\:
\hat{\cal U}[q_+](t_{f},t_{i}) \Bigr\rangle
_{\!\hat{\rho}_e^{\rm S}(t_i)}, 
\label{influence funct representation}
\ee
where we use 
$\langle \hspace{1.5ex} \rangle_{\!\hat{\rho}_e^{\rm S}(t_i)}$ to
denote an expectation value in the state described by 
$\hat{\rho}_e^{\rm \scriptscriptstyle S}(t_i)$.
From this expression, it follows that the influence
functional satisfies
\be
{\cal F}_{\rm IF}[q,q]=1, \hspace{7 ex} 
{\cal F}_{\rm IF}[q_-,q_+]=
{\cal F}_{\rm IF}^{\displaystyle\,\ast}[q_+,q_-], \hspace{7 ex}
\bigl|\hspace{0.2ex} 
{\cal F}_{\rm IF}[q_+,q_-]\hspace{0.2ex}\bigr|\leq 1, 
\label{influence funct properties}
\ee
or, equivalently, in terms of the influence action,
\be
S_{\rm IF}[q,q]=0, \hspace{7 ex} 
S_{\rm IF}[q_-,q_+]=
-S_{\rm IF}^{\displaystyle\,\ast}[q_+,q_-], \hspace{7 ex} 
{\rm Im}\, S_{\rm IF}[q_+,q_-] \geq 0,
\label{influence action properties}
\ee
and similar properties follow for $S_{\rm eff}[q_+,q_-]$.
A decoherence functional for the system, 
where the environment variables have been completely
integrated out, can now be introduced as the functional Fourier
transform of the CTP generating functional in the external 
sources:
\be
Z[j_+,j_-] \equiv 
\int\! {\cal D}[q_+]\;{\cal D}[q_-]\; 
{\cal D}[q_+,q_-] \;
e^{i \int\! dt\, (j_+ q_+ - j_- q_- )},
\label{decoherence functional}
\ee
that is, from (\ref{generating functional}) we have that
\be
{\cal D}[q_+,q_-]=
\rho_s (q_{+_{\scriptstyle i}},q_{-_{\scriptstyle i}};t_i ) \: 
\delta(q_{+_{\scriptstyle f}}\!-q_{-_{\scriptstyle f}}) \;
e^{{i \over \hbar}\, S_{\rm eff}[q_+,q_-]}.
\label{decoherence functional 2}
\ee
In the consistent histories approach to quantum mechanics,
${\cal D}[q_+,q_-]$ is known as the decoherence functional for
fine-grained histories of the system 
\cite{gell-mann-hartle,hartle,halliwell93,histories,dowker,halliwell}.

The environment of a system has to be understood as characterized by
all the quantum degrees of freedom which can affect the dynamics of the
system, but which are ``not accessible'' in the 
observations of that system. This environment includes in general 
an ``external'' environment (variables representing other particles,
or, in the context of field theory, other fields) and an 
``internal'' environment 
(some degrees of freedom which, from the fundamental quantum theory
point of view, would be associated to the same physical object as the
``system'' variables, but which are not directly probed in our
observations of the system) \cite{zurek,omnes}.
For instance, a problem which has been studied using the
influence functional method is that of quantum Brownian motion
\cite{feynman-vernon,feynman-hibbs,caldeira,%
hu-paz-zhang,hu-matacz94,hu-paz-zhang2,gell-mann-hartle,%
hartle,halliwell93,dowker,halliwell,hakim,grabert,brun}. 
In this problem, one is interested in the dynamics of a
macroscopic particle interacting with a medium
composed by a large number of other particles. In this example, 
one considers that the 
only ``observable'' system degree of freedom is the center of
mass position of the macroscopic particle, whereas the 
remaining microscopic degrees of freedom of the macroscopic
particle are considered as environmental variables.
Such ``internal'' environment degrees of freedom, and also those of
the particles of the medium (the ``external'' environment), 
are usually modelized as an infinite set of
harmonic oscillators. 
In the context of field theory, one
would typically consider as ``inaccessible'' to the observations the
modes of the field of interest with characteristic momenta higher than
some cut-off momentum \cite{lombardo,greiner,matacz}. 
In the case of the gravitational field, this has
been considered by Whelan \cite{whelan} in a toy model designed to
investigate the decoherence mechanism for gravity.

It is convenient at this stage to distinguish between these two kinds
of environmental variables, so let $Q$ represent the coordinates of
the ``external'' environment (the coordinates of ``other particles'') 
and $q_{\mbox{}_{\rm U}}$ the ``unobservable system'' coordinates (the
coordinates of the ``internal'' environment). As before, $q$ will
represent the ``true'' system coordinates. 
One could now simply replace $Q$ by $(Q,q_{\mbox{}_{\rm U}})$ 
in the previous expressions. However,
for convenience, we shall do 
the integrations in the environmental variables
in two steps. The action of the full system will be now written as
$S[q,q_{\mbox{}_{\rm U}},Q]$, and, as before, we shall assume a totally
factorizable initial density operator 
$\hat{\rho}^{\rm \scriptscriptstyle S}(t_i)=
\hat{\rho}_s^{\rm \scriptscriptstyle S}(t_i)\otimes 
\hat{\rho}_{\mbox{}_{\rm U}}^{\rm \scriptscriptstyle S}(t_i) 
 \otimes \hat{\rho}_e^{\rm \scriptscriptstyle S}(t_i)$, 
which leads to an initial density matrix in coordinate representation 
of the form 
$\rho(q,q_{\mbox{}_{\rm U}},Q;
q^{\prime},q_{\mbox{}_{\rm U}}^{\prime},Q^{\prime};t_i)=
\rho_s(q,q^{\prime};t_i)\,
\rho_{\mbox{}_{\rm U}}(q_{\mbox{}_{\rm U}},
              q_{\mbox{}_{\rm U}}^{\prime};t_i)\,
\rho_e(Q,Q^{\prime};t_i)$ (notice that we are now using the subindex 
$e$ for the ``external'' environment).
Such a factorization is based on the assumption that the interactions
between the three subsystems can be neglected for times
$t \leq t_{i}$. Unfortunately, in most situations, this assumption 
does not seem to be very physically reasonable, especially for the
``true'' system-``internal'' environment interactions. 
One would need to consider the generalization of the
formalism to a non-factorizable initial density operator 
mentioned above and the analysis would be more complicated.
We start defining 
\bea
&& \hspace{-10ex}
e^{{i \over \hbar}\,\left(\, S_{s}^{\rm eff}[q_+]-S_{s}^{\rm eff}[q_-] 
+S_{se}^{\rm eff}[q_+,Q_+;q_-,Q_-] \,\right)} \nn  \\
&& \hspace{-5ex} \equiv 
\int\! {\cal D}[q_{\mbox{}_{\rm U}+}]\;{\cal D}[q_{\mbox{}_{\rm U}-}]\;
\rho_{\mbox{}_{\rm U}} (q_{{\mbox{}_{\rm U}+}_{\scriptstyle i}},
         q_{{\mbox{}_{\rm U}-}_{\scriptstyle i}};t_i ) \:
\delta(q_{{\mbox{}_{\rm U}+}_{\!\scriptstyle f}}\!-
q_{{\mbox{}_{\rm U}-}_{\!\scriptstyle f}}) \;
e^{{i \over \hbar}\, \left(\,S[q_+,q_{\mbox{}_{\rm U}+},Q_+]
-S[q_-,q_{\mbox{}_{\rm U}-},Q_-]\, \right)},
\label{s-e effective actions}
\eea
where the effective action for the system $S_{s}^{\rm eff}[q]$ is
chosen to be real and local. Notice that the effective action
$S_{se}^{\rm eff}[q_+,Q_+;q_-,Q_-]$ has analogous properties to those
of $S_{\rm IF}$ in (\ref{influence action properties}).
We introduce now an effective influence functional and an
effective influence action as
\be
{\cal F}^{\rm eff}_{\rm IF}[q_+,q_-] \equiv  e^{{i \over \hbar}\,
  S^{\rm eff}_{\rm IF}[q_+,q_-]}  
\equiv \hspace{-0.1ex}
\int\! {\cal D}[Q_+]\;{\cal D}[Q_-]\;
\rho_e (Q_{+_{\scriptstyle i}},Q_{-_{\scriptstyle i}};t_i ) \:
\delta(Q_{+_{\scriptstyle f}}\!-Q_{-_{\scriptstyle f}}) \;
e^{{i \over \hbar}\, S_{se}^{\rm eff}[q_+,Q_+;q_-,Q_-]}.
\label{effective influence functional}
\ee
With these definitions, the effective action of Feynman and Vernon,
$S_{\rm eff}[q_+,q_-]$, 
which appears in expression (\ref{generating functional}) can be
written as
\be
S_{\rm eff}[q_+,q_-]\equiv S_s^{\rm eff}[q_+]-S_s^{\rm eff}[q_-]
+S^{\rm eff}_{\rm IF}[q_+,q_-].
\label{effective action 2}
\ee
Note that, since $S_{\rm eff}[q_+,q_-]$ satisfies the same properties
as $S_{\rm IF}$ in (\ref{influence action properties}), it follows
from the last expression that $S^{\rm eff}_{\rm IF}$ has also these 
properties.


\subsection{\hspace{-2.5ex}. The ``naive'' semiclassical 
approximation}
\label{subsec:naive semiclassical}


The usual ``naive'' semiclassical approximation for the system
variables consists in performing a ``tree level'' approximation in the
path integrals involving the $q$ variables in expression 
(\ref{generating functional})
\cite{calzettahu,greiner,cv96,ccv97,campos-hu,shaisultanov}. 
Therefore, the CTP generating functional
is approximated by
\be
Z[j_+,j_-] \simeq e^{{i \over \hbar}\, 
\left(S_{\rm eff}\bigl[\bar{q}_+^{\scriptscriptstyle (0)}
\hspace{-0.2ex}[j]\, , \,
\bar{q}_-^{\scriptscriptstyle (0)}\hspace{-0.2ex}[j]\bigr]+ 
 \hbar \int\! dt\,  j_{\mbox{}_+} 
\bar{q}_+^{\scriptscriptstyle (0)}
\hspace{-0.2ex}[j]-
 \hbar \int\! dt\,  j_{\mbox{}_-} 
\bar{q}_-^{\scriptscriptstyle (0)}
\hspace{-0.2ex}[j]
\right)},
\label{semiclass approx}
\ee
where 
$\bar{q}_{\pm}^{\scriptscriptstyle (0)}\hspace{-0.2ex}[j] 
\!\equiv\! 
\bar{q}_{\pm}^{\scriptscriptstyle (0)}\hspace{-0.2ex}[j_+,j_-]$ are
solutions of the classical equations of motion for the action 
$S_{\rm eff}[q_+,q_-]+
\hbar \int\! dt\, j_+ q_+ -\hbar \int\! dt\, j_- q_-$, that is,
\be
{\delta S_{\rm eff}[\bar{q}_+^{\scriptscriptstyle (0)} , 
\bar{q}_-^{\scriptscriptstyle (0)}] \over \delta q_{\pm}(t)}
= \mp \, \hbar j_{\pm}(t),
\label{semiclass eqs with j's}
\ee
which satisfy the boundary condition 
$\bar{q}_+^{\scriptscriptstyle (0)}(t_f)
\!=\!\bar{q}_-^{\scriptscriptstyle (0)}(t_f)$. 
Whenever this approximation is valid, we can see from 
(\ref{semiclass approx}), (\ref{semiclass eqs with j's}) and 
(\ref{green functions}) that 
$\langle \hat{q}(t) \rangle \simeq q^{\scriptscriptstyle (0)}(t)$, 
with $q^{\scriptscriptstyle (0)} \equiv  
\bar{q}_{+}^{\scriptscriptstyle (0)}\hspace{-0.2ex}
[j_+\!=\!j_-\!=\!0]=
\bar{q}_{-}^{\scriptscriptstyle (0)}\hspace{-0.2ex}
[j_+\!=\!j_-\!=\!0]$, that is, $q^{\scriptscriptstyle (0)}(t)$ is a
solution of the two equivalent equations:
\be
\left. {\delta S_{\rm eff}[q_+,q_-] \over \delta q_+(t)}
\right|_{q_+=q_-=q^{\scriptscriptstyle (0)}} =0,
\hspace{10ex}
\left. {\delta S_{\rm eff}[q_+,q_-] \over \delta q_-(t)}
\right|_{q_+=q_-=q^{\scriptscriptstyle (0)}} =0.
\label{semiclassical eq}
\ee
One can see that these two equations are actually the same
equation, and that this equation is real.
This is the semiclassical equation for the system variables.
In a naive way, one would think that, when the 
above semiclassical
approximation is valid, the system would behave as a classical system 
described by the coordinate functions $q^{\scriptscriptstyle (0)}(t)$,
{\it i.e.}, that one could substitute the description of the system in
terms of the operators $\hat{q}(t)$ by a classical description in terms
of the functions $q^{\scriptscriptstyle (0)}(t)$. However, one can see
from (\ref{semiclass approx}), (\ref{semiclass eqs with j's})
and (\ref{green functions}) that,
in general,
\be
\left\langle\, \tilde{\rm T}[\hat{q}(t_1^{\prime}) \cdots 
\hat{q}(t_s^{\prime})] \,{\rm T}[\hat{q}(t_1) \cdots
\hat{q}(t_r)]\, \right\rangle \simeq \hspace{-2ex}/
\hspace{1ex}
q^{\scriptscriptstyle (0)}(t_1) \cdots
q^{\scriptscriptstyle (0)}(t_r)\, 
q^{\scriptscriptstyle (0)}(t_1^{\prime}) \cdots 
q^{\scriptscriptstyle (0)}(t_s^{\prime}).
\ee
Thus, in general, whenever the above approximations are valid, we can 
only interpret the solutions of the semiclassical equation as
representing the expectation value of the operators $\hat{q}(t)$.


\subsection{\hspace{-2.5ex}. Further coarse-graining and 
decoherence}
\label{subsec:decoherence}


Decoherence takes place in a set of quantum-mechanical variables
when the quantum interference effects are (in general,
approximately) suppressed in the 
description of the properties of a physical system which are 
associated to that variables.
When this happens, such decoherent variables
can be described in an effective probabilistic way. 
In the Heisenberg picture, we will say that  
a set of variables decohere when the description in terms of the
operators corresponding to these variables can be replaced by 
an effective description in terms of a set of
classical random variables, 
in the sense that the quantum Green functions for such operators 
become approximately equal to the moments of the classical random
variables. For the Green functions (\ref{green functions}),
it is easy to see that this would hold in an exact way if the CTP
generating functional (\ref{generating functional}) depended on the
sources $j_{\pm}$ only as a functional $\Phi_q[j_+\!-\!j_-]$ of the
difference $j_+ - j_-$, or, equivalently, if the 
decoherence functional (\ref{decoherence functional}) could be 
written as
${\cal D}[q_+,q_-]= {\cal P}_q[q_+] \, \delta[q_+-q_-]$.
However, in practice, one finds that this condition is usually too
strong to be satisfied, even in an approximate way
\cite{gell-mann-hartle,hartle,halliwell93,histories,dowker,%
halliwell,whelan}.
One needs to introduce further coarse-graining in the system degrees
of freedom in order to achieve decoherence.
Let us then introduce coarse-grained system operators, which 
correspond to imprecisely specified values of the system
coordinates. In the Heisenberg picture, such operators can be
defined as 
\be
\hat{q}_c(t) \equiv \sum_{\bar{q}} \bar{q} \:\hat{P}_{\bar{q}}(t),
\ee 
where $\hat{P}_{\bar{q}}(t)$ is a set of projection operators,
labeled by some variables $\bar{q}$
(these are often discrete variables), of the form
\be
\hat{P}_{\bar{q}}(t) = \int\! dq \, dq_{\mbox{}_{\rm U}}
\hspace{0.1ex} dQ \: \gamma(q-\bar{q}) \: 
|q,q_{\mbox{}_{\rm U}} ,Q,t \rangle
\langle q,q_{\mbox{}_{\rm U}} ,Q,t|.
\ee
Here $\left\{  |q,q_{\mbox{}_{\rm U}} ,Q,t \rangle \right\}$
is the basis of
eigenstates of the operators $\hat{q}(t)$, 
$\hat{ q}_{\mbox{}_{\rm U}} (t)$ and $\hat{Q}(t)$, and 
$\gamma$ is a real function. 
We shall assume coarse-grainings of characteristic sizes $\sigma$,
that is, such that the function $\gamma(q-\bar{q})$ vanishes or has
negligible values for $q$ outside a cell $I_{\bar{q}}$ of sizes 
$\sigma$ centered around  $\bar{q}$. This means that
\be
\int\! dq \;\gamma(q-\bar{q}) \: f(q) \simeq 
\int_{I_{\bar{q}}} \! dq \;\gamma(q-\bar{q}) \: f(q),
\label{c-g characteristic sizes}
\ee
for any function $f(q)$. In addition, the function $\gamma$ must be
chosen in such a way that the set
of projection operators is (at least, approximately)
exhaustive and mutually exclusive,
which means that
\be
\sum_{\bar{q}}  \hat{P}_{\bar{q}}(t)= \hat{I},
\hspace{10ex}
\hat{P}_{\bar{q}}(t) \hat{P}_{\bar{q}^\prime}(t)= 
\delta_{\bar{q} \bar{q}^\prime}\, \hat{P}_{\bar{q}}(t),
\label{proj properties}
\ee
where $\hat{I}$ is the identity operator. For specific examples of
operators satisfying the above properties in an exact or in an
approximate way, see Refs.~\cite{dowker,halliwell}.

Next, we can introduce a family of 
decoherence functions for coarse-grained
histories of the system 
\cite{gell-mann-hartle,hartle,halliwell93,histories,paz-zurek,%
dowker,halliwell}.
In order to do so, let us consider a set $\{t_1, \dots , t_N \}$ of
$N$ instants of time, such that 
$t_k < t_{k+1}$, $k = 0, \dots , N$, with $t_0 \equiv t_i$ and 
$t_{N+1} \equiv t_f$.
Introducing two sets of values of
$\bar{q}$ associated to such set of instants, 
$\{ \bar{q}_+ \} \equiv \{ \bar{q}_{+_1}, \dots , \bar{q}_{+_N}\}$
and 
$\{ \bar{q}_- \} \equiv \{ \bar{q}_{-_1}, \dots , \bar{q}_{-_N}\}$,
the decoherence function for this pair of
``coarse-grained histories'' of the system is defined as
\be
{\cal D}_c( \{ \bar{q}_+ \},\{ \bar{q}_-\} )_{(t_1, \dots , t_N)}
\equiv
{\rm Tr} \! \left[
\hat{P}_{\bar{q}_{+ _N}}(t_N) \cdots \hat{P}_{\bar{q}_{+ _1}}(t_1)
\, \hat{\rho} \, \hat{P}_{\bar{q}_{- _1}}(t_1) \cdots 
\hat{P}_{\bar{q}_{- _N}}(t_N)
\right], 
\label{c-g decoh funct}
\ee
where $\hat{\rho}$ is the density operator describing the
state of the entire system (system plus environment)
in the Heisenberg picture (${\cal D}_c$ is often called
decoherence ``functional'' in the literature, but, for each
set $\{t_1, \dots , t_N \}$, this is actually a function of $2N$
variables).
These decoherence functions
can be written in a path integral form as
\be
{\cal D}_c( \{ \bar{q}_+ \},\{ \bar{q}_-\} )_{(t_1, \dots , t_N)} = 
\! \int\! {\cal D}[q_+]\,{\cal D}[q_-]\,
\prod_{k=1}^N 
\gamma (q_+(t_k) \!-\! \bar{q}_{+_k})\, 
\gamma (q_-(t_k) \!-\! \bar{q}_{-_k})\:
{\cal D}[q_+,q_-], 
\label{c-g decoh funct 2}
\ee
where ${\cal D}[q_+,q_-]$ is the decoherence functional for
fine-grained histories of the system (\ref{decoherence functional}).
From the definition (\ref{c-g decoh funct}) and the properties
(\ref{proj properties}), one can show that these decoherence functions
have the properties
\be
\sum_{\{ \bar{q}_+ \}} \sum_{\{ \bar{q}_- \}}
{\cal D}_c( \{ \bar{q}_+ \},\{ \bar{q}_-\} ) = 1,
\hspace{10ex}
{\cal D}_c( \{ \bar{q}_- \},\{ \bar{q}_+\} ) =
{\cal D}_c^{\displaystyle \ast}( \{ \bar{q}_+ \},\{ \bar{q}_-\} ),
\label{decoh funct properties}
\ee 
and that the diagonal elements of the
decoherence functions (the values of those functions in
the limit $\bar{q}_{-_k}\!\rightarrow \! \bar{q}_{+_k}$) are
positive. 
For $N \!>\! 1$, we can also see that,
if we divide the set
$\{t_1, \dots , t_N \}$ into a subset of $M \!<\! N$ instants,
$\{t_1^{\prime}, \dots , t_M^{\prime} \} 
\!\subset\! \{t_1, \dots , t_N \}$, with 
$t_1^{\prime} < \cdots <t_M^{\prime}$, and the subset of the remaining 
$L \!\equiv\! M\!-\!N$ instants, denoted as 
$\{t_1^{\prime \prime}, \dots , t_L^{\prime \prime} \}$
[{\it i.e.}, 
$\{t_1, \dots , t_N \}\!=\! \{t_1^{\prime}, \dots , t_M^{\prime} \} 
\cup \{t_1^{\prime \prime}, \dots , t_L^{\prime \prime} \}$], then
\be
{\cal D}_c( \{ \bar{q}_+ \}_{\!\mbox{}_{M}},
\{ \bar{q}_-\}_{\!\mbox{}_{M}} )_{(t_1^{\prime}, \dots , t_M^{\prime})}
= \sum_{ \{ \bar{q}_+ \}_{\!\mbox{}_{L}} }  
\sum_{\{ \bar{q}_- \}_{\!\mbox{}_{L}} } 
{\cal D}_c( \{ \bar{q}_+ \}_{\!\mbox{}_{N}},
\{ \bar{q}_-\}_{\!\mbox{}_{N}} )_{(t_1, \dots , t_N)},
\label{decoh funct prop 4}
\ee
with
$\{ \bar{q}_{\pm} \}_{\!\mbox{}_{M}} \!\equiv\!
\{ \bar{q}_{\pm}(t_1^{\prime}), \dots, 
\bar{q}_{\pm}(t_M^{\prime}) \}$,
$\{ \bar{q}_{\pm} \}_{\!\mbox{}_{L}} \!\equiv\!
\{ \bar{q}_{\pm}(t_1^{\prime\prime}), \dots, 
\bar{q}_{\pm}(t_L^{\prime\prime}) \}$,
where we use the notation 
$\bar{q}_{\pm}(t_k) \!\equiv\! \bar{q}_{\pm_k}$, for 
$k \!=\! 1, \dots, N$,
and
$\{ \bar{q}_{\pm} \}_{\!\mbox{}_{N}} \!\equiv\!
\{ \bar{q}_{{\pm}_1}, \dots, \bar{q}_{{\pm}_N} \}$.

To make contact with the CTP formalism, let us introduce
now, in analogy with (\ref{decoherence functional}), 
a family of 
generating functions for the coarse-grained system degrees of
freedom as the following Fourier series:
\be
Z_c( \{ j_+ \},\{ j_- \} )_{(t_1, \dots , t_N)} 
\equiv 
\sum_{ \{ \bar{q}_+ \}} \sum_{ \{ \bar{q}_- \}}
{\cal D}_c( \{ \bar{q}_+ \},\{ \bar{q}_-\} )_{(t_1, \dots , t_N)}\;
e^{i \sum_{k=1}^{N}
(j_{+_k} \bar{q}_{+_k} - j_{-_k} \bar{q}_ {-_k})} ,
\label{c-g generating funct}
\ee
where 
$\{ j_{\pm} \} \equiv \{ j_{\pm_1}, \dots , j_{\pm_N}\}$.
Note that the properties (\ref{decoh funct properties}) for the
decoherence functions are equivalent to
\be
Z_c( \{0 \},\{ 0 \} )=1, 
\hspace{10ex}
Z_c( \{ j_- \},\{ j_+ \} )
=Z_c^{\,\displaystyle \ast}( \{ j_+ \},\{ j_- \} ).
\label{c-g generating funct properties}
\ee 
From the generating function (\ref{c-g generating funct}),
we can compute the Green functions 
\[
G_{c \; m_1 \cdots \, m_s}^{n_1 \cdots \, n_r}
(t_1^{\prime}, \dots, t_r^{\prime};
t_1^{\prime \prime}, \dots, t_s^{\prime \prime})
\equiv
\left\langle\, \tilde{\rm T}[\hat{q}_c^{m_1}(t_1^{\prime \prime}) 
\cdots
\hat{q}_c^{m_s}(t_s^{\prime \prime})] \,
{\rm T}[\hat{q}_c^{n_1}(t_1^{\prime}) \cdots
\hat{q}_c^{n_r}(t_r^{\prime})]\, \right\rangle,
\]
with $n_1, \dots, n_r, m_1, \dots, m_s \!\in \! 
{\rm I\hspace{-0.4 ex}N}$,
$\{t_1^{\prime}, \dots , t_r^{\prime} \}  \!\subseteq \!
\{t_1, \dots , t_N \}$ and
$\{t_1^{\prime \prime}, \dots , t_s^{\prime \prime} \}  \!\subseteq \!
\{t_1, \dots , t_N \}$ (thus, $r,s \leq N$):
\be
G_{c \; m_1 \cdots \, m_s}^{n_1 \cdots \, n_r} \!\hspace{0.1ex}
(t_1^{\prime}, \dots, t_r^{\prime};
t_1^{\prime \prime}, \dots, t_s^{\prime \prime})= \!
\left. { \left(-i \partial \right)
^{n_1+ \cdots +n_r+ m_1+ \cdots +m_s} 
Z_c( \{ j_+ \},\{ j_- \} )_{(t_1, \dots , t_N)}
\over 
\left[ \partial j_+(t_1^{\prime}) \right]^{\hspace{-0.1 ex} n_1} 
\!\! \cdots \!
\left[ \partial j_+(t_r^{\prime}) \right]^{\hspace{-0.1 ex} n_r}
\!
\left[-\partial j_-(t_1^{\prime \prime}) 
\right]^{\hspace{-0.1 ex} m_1}
\!\! \cdots
\!\left[- \partial j_-(t_s^{\prime \prime})
\right]^{\hspace{-0.1 ex} m_s} \!}
 \right|_{\{j_\pm \}=\{0\} } 
\!,
\label{c-g green funct}
\ee
where $j_\pm (t_k) \!\equiv\! j_{\pm_k}$, for $k \!=\! 1, \dots, N$.
The property (\ref{decoh funct prop 4}) can also be written
in terms of the corresponding generating functions as
\be
Z_c(\{ j_+ \}_{\!\mbox{}_{M}},
\{ j_-\}_{\!\mbox{}_{M}} )_{(t_1^{\prime}, \dots , t_M^{\prime})}
=\left.
Z_c( \{ j_+ \}_{\!\mbox{}_{N}},
\{ j_- \}_{\!\mbox{}_{N}} )_{(t_1, \dots , t_N)}
\right|_{\{j_\pm \}_{\!\mbox{}_{L}}=\{0\} },
\label{c-g generating funct prop 4}
\ee
with the notation
$\{ j_{\pm} \}_{\!\mbox{}_{M}} \!\equiv\!
\{ j_{\pm}(t_1^{\prime}), \dots, 
j_{\pm}(t_M^{\prime}) \}$, and similarly
for $\{ j_{\pm} \}_{\!\mbox{}_{L}}$ and
$\{ j_{\pm} \}_{\!\mbox{}_{N}}$. 
Notice that this last property is consistent with 
(\ref{c-g green funct}), in the sense that, for instance, 
$G_c^{n_1 n_2}(t_1^{\prime},t_2^{\prime})$
can be equally computed either from
$Z_c(\{ j_+ \}_{\mbox{}_{2}},
\{ j_-\}_{\mbox{}_{2}} )_{(t_1^{\prime},t_2^{\prime})}$,
or from
$Z_c( \{ j_+ \}_{\!\mbox{}_{N}},
\{ j_- \}_{\!\mbox{}_{N}} )_{(t_1, \dots , t_N)}$, with
$N>2$.

Having introduced the coarse-grained description of the system in
terms of the operators $\hat{q}_c(t)$, we can now sketch the
decoherence mechanism for them. For the Green functions
(\ref{c-g green funct}), one can show that 
the decoherence condition described
above holds in an exact way if the 
generating function (\ref{c-g generating funct}) depends on the
sources $j_{\pm_k}$ only as a function of the
differences $j_{+_k}\!-\!j_{-_k}$, {\it i.e.}, as
$\Phi_{\bar{q}}( \{ j_+\!-\!j_- \} )_{(t_1, \dots , t_N)}$.
Then, introducing the
Fourier series corresponding to $\Phi_{\bar{q}}$, we can write
\be
Z_c( \{ j_+ \},\{ j_- \})_{(t_1, \dots , t_N)}=
\Phi_{\bar{q}}( \{ j_+ \hspace{-0.2ex}-\hspace{-0.2ex} j_- \} )
_{(t_1, \dots , t_N)}
\equiv \sum_{\{ \bar{q}\} } 
{\cal P}_{\bar{q}}( \{ \bar{q} \} )_{(t_1, \dots , t_N)} \; 
e^{i \sum_{k=1}^{N} 
\bar{q}_k  ( j_{+_k} - j_{-_k}) }.
\label{decoherence condition}
\ee
Note from the last expression that, if we interpret the function
${\cal P}_{\bar{q}}$ as the probability distribution for a set of
random 
variables $\bar{q}_k$, $k \!=\! 1, \dots, N$, associated to the
instants $t_k$, then $\Phi_{\bar{q}}$ is the 
corresponding characteristic function. Therefore, from 
(\ref{c-g green funct}), we get
\bea
&&\hspace{-4ex}
G_{c \; m_1 \cdots \, m_s}^{n_1 \cdots \, n_r} 
(t_1^{\prime}, \dots, t_r^{\prime};
t_1^{\prime \prime}, \dots, t_s^{\prime \prime})=
\left. 
{ \left(-i \partial \right)
^{n_1+ \cdots +n_r+ m_1+ \cdots +m_s} 
\Phi_{\bar{q}}( \{ j \} )_{(t_1, \dots , t_N)} 
\over 
\left[ \partial j(t_1^{\prime}) \right]^{n_1} 
\cdots 
\left[ \partial j(t_r^{\prime}) \right]^{n_r}
\left[\partial j(t_1^{\prime \prime}) 
\right]^{m_1}
\cdots
\left[\partial j(t_s^{\prime \prime})
\right]^{m_s} }
\right|_{\{j \}=\{0\} }  \nn  \\
&&\hspace{-4ex}
=\!\hspace{-0.1ex} \sum_{\{ \bar{q}\} } \hspace{-0.1ex}
{\cal P}_{\bar{q}}(\hspace{-0.1ex} \{ \bar{q} \} \hspace{-0.1ex} )
_{\hspace{-0.1ex}(t_1, \dots , t_N)} \, 
\bar{q}^{n_1 \!}(t_1^{\prime}) 
\hspace{-0.2ex} \cdots \hspace{-0.2ex}  
\bar{q}^{n_r \!}(t_r^{\prime})\,
\bar{q}^{m_1 \!}(t_1^{\prime \prime}) 
\hspace{-0.2ex} \cdots \hspace{-0.2ex}
\bar{q}^{m_s \!}(t_s^{\prime \prime}) 
\hspace{-0.2ex} \equiv \hspace{-0.2ex}
\left\langle \bar{q}^{n_1 \!}(t_1^{\prime}) 
\hspace{-0.2ex} \cdots \hspace{-0.2ex}
\bar{q}^{n_r \!}(t_r^{\prime})\,
\bar{q}^{m_1 \!}(t_1^{\prime \prime}) 
\hspace{-0.2ex} \cdots \hspace{-0.2ex}
\bar{q}^{m_s \!}(t_s^{\prime \prime}) 
\right\rangle_{\hspace{-0.2ex} c} \hspace{-0.2ex}, \nn  \\
\mbox{}
\label{correlation functions}
\eea
where $\langle \hspace{1.5ex} \rangle_c$ means statistical average
of the random variables, and we use the notation
$\bar{q}(t_k) \!\equiv\! \bar{q}_k$,
$j(t_k) \!\equiv\! j_k$, for $k \!=\! 1, \dots, N$.
Note that, if (\ref{decoherence condition}) is satisfied,
then the property (\ref{c-g generating funct prop 4}) reduces to
\be
\Phi_{\bar{q}}( \{ j \}_{\!\mbox{}_{M}} )
_{(t_1^{\prime}, \dots , t_M^{\prime})}
= \left.
\Phi_{\bar{q}}( \{ j \}_{\!\mbox{}_{N}} )_{(t_1, \dots , t_N)} 
\right|_{\{ j \}_{\!\mbox{}_{L}}=\{0\} },
\label{prop 4}
\ee
or, equivalently,
\be
{\cal P}_{\bar{q}}( \{ \bar{q} \}_{\!\mbox{}_{M}} )
_{(t_1^{\prime}, \dots , t_M^{\prime})}
= \sum_{ \{ \bar{q} \}_{\!\mbox{}_{L}}}
{\cal P}_{\bar{q}}( \{ \bar{q} \}_{\!\mbox{}_{N}} )
_{(t_1, \dots , t_N)}.
\label{prop 4 bis}
\ee
This last property is a necessary condition for the probabilistic
interpretation (\ref{correlation functions}) to be consistent.

The conditions for decoherence (\ref{decoherence condition}) can be
written in terms of the corresponding decoherence functions as
\be
{\cal D}_c( \{ \bar{q}_+ \},\{ \bar{q}_-\} )_{(t_1, \dots , t_N)} 
={\cal P}_{\bar{q}}( \{ \bar{q}_+ \} )_{(t_1, \dots , t_N)} 
\prod_{k=1}^N \delta_{\bar{q}_{+_k} \bar{q}_{-_k}}.
\label{decoherence condition 2}
\ee
These are actually the conditions for decoherence
of coarse-grained system variables  
as stated in the consistent histories
formulation of quantum mechanics 
\cite{gell-mann-hartle,hartle,halliwell93,histories,paz-zurek,%
dowker,halliwell}.
Notice, from (\ref{proj properties}), that 
(\ref{decoherence condition 2}) is always satisfied for a single
instant of time ({\it i.e.}, when $N \!=\! 1$) \cite{dowker}.

We can now check that the interpretation of 
${\cal P}_{\bar{q}}$ as a probability function
is actually correct. 
From the second of the properties (\ref{decoh funct properties}),
we have that 
${\cal P}_{\bar{q}}^{\displaystyle \ast}( \{ \bar{q} \} )
={\cal P}_{\bar{q}}( \{ \bar{q} \} )$, {\it i.e.},
${\cal P}_{\bar{q}}$ is real.
Since the diagonal elements of the
decoherence functions are positive,  
${\cal P}_{\bar{q}}( \{ \bar{q} \} )$ is also positive.
These two properties of 
${\cal P}_{\bar{q}}( \{ \bar{q} \} )_{(t_1, \dots , t_N)}$,
together with (\ref{prop 4 bis}),
are enough to
guarantee that it can be properly interpreted as the probability
distribution for a set of random variables associated to the instants
$t_1, \dots , t_N$. 
From the first of
the relations (\ref{decoh funct properties}), which yields
$\sum_{\{ \bar{q}\} } {\cal P}_{\bar{q}}( \{ \bar{q} \} )=1$, 
it follows that this probability distribution is normalized.

In practice, the conditions for decoherence described above will be
usually only satisfied in an approximate way. 
Approximate decoherence is typically achieved through a mechanism
which was proposed by Gell-Mann and Hartle 
\cite{gell-mann-hartle,hartle}. 
To see how this works, note that, if we assume
coarse-grainings of characteristic sizes $\sigma$
[see (\ref{c-g characteristic sizes})], 
and using (\ref{decoherence functional 2}),
we can write the decoherence 
function (\ref{c-g decoh funct 2}) as
\bea
&&\hspace{-3.9ex}
{\cal D}_c( \{ \bar{q}_+ \},\{ \bar{q}_-\} )_{(t_1, \dots , t_N)}
\! \simeq \! \int_{ \!
\mbox{}_{ \scriptstyle
\{ I_{\bar{q}_{+}} \}, \{ I_{\bar{q}_{-}} \} }}
\hspace{-3ex}
{\cal D}[q_+^{(0)}]\,{\cal D}[q_-^{(0)}]
\prod_{k=1}^N {\cal D}[q_+^{(k)}]\,{\cal D}[q_-^{(k)}]\,
\rho_s 
\bigl(q^{(0)}_{+_{\scriptstyle i}},q^{(0)}_{-_{\scriptstyle i}}
;t_i \bigr)
\, \delta 
\bigl(q^{(N)}_{+_{\scriptstyle f}}\!-q^{(N)}_{-_{\scriptstyle f}} 
\bigr)  
\nn \\
&&\hspace{-4ex} \times 
\delta 
\Bigl(q^{\hspace{-0.1ex}(k-1)}_+\hspace{-0.1ex}(t_k) \!-\!  
q^{\hspace{-0.1ex}(k)}_+\hspace{-0.1ex}(t_k)\Bigr) 
\hspace{0.2ex}
\delta
\Bigl(q^{\hspace{-0.1ex}(k-1)}_-\hspace{-0.1ex}(t_k) \!-\!  
q^{\hspace{-0.1ex}(k)}_-\hspace{-0.1ex}(t_k)\Bigr)
\hspace{0.2ex} 
\gamma (q^{\hspace{-0.1ex}(k)}_+\hspace{-0.1ex}(t_k) 
\!-\! \bar{q}_{+_k}) 
\hspace{0.2ex}
\gamma (q^{\hspace{-0.1ex}(k)}_-\hspace{-0.1ex}(t_k) 
\!-\! \bar{q}_{-_k}) 
\prod_{k=0}^N \!
e^{{i \over \hbar}\, S_{\rm eff}[q_+^{\hspace{-0.1ex}(k)}
\hspace{-0.4ex},q_-^{\hspace{-0.1ex}(k)}]},
\nn \\
\mbox{}
\label{c-g decoh funct 3}
\eea
where each path integration $\int {\cal D}[q_{\pm}^{(k)}]$, for
$k=0, \dots, N$, is over paths $q_{\pm}^{(k)}(t)$ with
$t \in [t_k,t_{k+1}]$, being $t_0 \equiv t_i$ and 
$t_{N+1} \equiv t_f$, and we have used a notation to indicate that 
these paths are restricted to pass through the cells 
$I_{\bar{q}_{\pm _k}}$ at the instants $t_k$, 
for $k=1, \dots, N$.
From (\ref{effective action 2}), the modulus of each factor
$\exp \bigr(
{{i \over \hbar}\hspace{0.2ex} S_{\rm eff}[q_+^{\hspace{-0.1ex}(k)}
,q_-^{\hspace{-0.1ex}(k)}]} \bigl)$ 
in the last expression is 
$\exp \bigr(
{-{1 \over \hbar} \hspace{0.2ex} {\rm Im}\, S_{\rm IF}^{\rm eff}
[q_+^{\hspace{-0.1ex}(k)} ,q_-^{\hspace{-0.1ex}(k)}]} \bigl)$.
Then, if for every $k=0, \dots, N$, 
${\rm Im}\, S_{\rm IF}^{\rm eff}
[q_+^{\hspace{-0.1ex}(k)} ,q_-^{\hspace{-0.1ex}(k)}]$, 
which is always positive or zero, 
is much larger than $\hbar$ whenever the
differences $|q_+^{\hspace{-0.1ex}(k)}\!-q_-^{\hspace{-0.1ex}(k)}|$ 
are larger than some ``cut-off'' sizes
$d^{(k)}$, 
the integrand in (\ref{c-g decoh funct 3}) will be only
non-negligible for 
$|q_+^{\hspace{-0.1ex}(k)}\!-q_-^{\hspace{-0.1ex}(k)}| \leq 
d^{(k)}$.
If the characteristic sizes $\sigma$ of the coarse-graining
satisfy $\sigma \!\gg \!d^{(k)}$, then the ``off-diagonal'' elements
of ${\cal D}_c( \{ \bar{q}_+ \},\{ \bar{q}_-\} )_{(t_1, \dots , t_N)}$
are negligible and one has approximate decoherence 
\cite{gell-mann-hartle,hartle}.
We should stress that
$S_{\rm IF}^{\rm eff}[q_+,q_-]$ is the result of integrating out both
the ``external'' environment degrees of freedom and also the system
degrees of freedom which are ``not accessible'' to the observations
(the ``internal'' environment). In general, these
two integrations play an important role in the achievement of this
sufficient condition for approximate decoherence. 
A characterization of the degree of approximate
decoherence has been given in Ref.~\cite{dowker} (see also 
Refs.~\cite{histories,halliwell93}).

Typically, $d^{(k)}$ can be estimated in terms of 
$\Delta t_k \!\equiv \! t_{k+1} \!-\! t_k$.
When this is the case, one usually finds that the Gell-Mann and Hartle
mechanism for approximate decoherence works provided all the time
intervals satisfy $\Delta t_k \!\geq\! \Delta t_c$, 
$k=0, \dots, N$, where $\Delta t_c$ is sufficiently larger than some
characteristic decoherence time scale 
$t_{\scriptscriptstyle \! D}$ ($t_{\scriptscriptstyle \! D}$ 
can be written
in terms of $\sigma$ and some parameters characterizing the
environment and the system-environment couplings)
\cite{gell-mann-hartle,hartle,paz-zurek}.
For $\Delta t_c$ one should take the smallest value compatible with a
specified degree of approximate decoherence. In this sense, we can
think of a coarse-graining as characterized both by the sizes
$\sigma$ and by the time scale $\Delta t_c$.


\subsection{\hspace{-2.5ex}. Effective equations of motion 
for the system} 
\label{subsec:effective eqs}


Assuming that the mechanism for approximate decoherence 
described in the previous subsection works, an approximate effective
description of the coarse-grained system variables in terms of a set
of random variables [in the sense of 
Eq.~(\ref{correlation functions})] is available, at least for instants
of time satisfying $\Delta t_k \!\geq\! \Delta t_c$, for
$k=0, \dots, N$. The corresponding probability distribution
${\cal P}_{\bar{q}}( \{ \bar{q} \} )_{(t_1, \dots , t_N)}$ is given
by the diagonal elements of the decoherence function 
(\ref{c-g decoh funct}).
We shall next make an estimation of this probability distribution.  
This follows essentially the derivation of Gell-Mann and Hartle in 
Refs.~\cite{gell-mann-hartle,hartle}. 
For alternative derivations for more specific models, 
see Refs.~\cite{halliwell93,dowker,halliwell}.
Introducing the new variables
$q_{\scriptscriptstyle \Delta}\!\equiv\! q_+\!-q_-$ and 
$q_{\scriptscriptstyle \Sigma}\!\equiv\! {1 \over 2}\, (q_+\!+\!q_-)$,
and similarly for $\bar{q}_{\pm_k}$, 
and assuming that $\sigma \gg d^{(k)}$, note first, from 
(\ref{c-g decoh funct 3}), that 
the restrictions on the integration over 
$q_{\scriptscriptstyle \Delta}$ coming from the coarse-graining can be 
neglected in the diagonal elements of this decoherence function.
Therefore, using (\ref{c-g decoh funct 2}) and
(\ref{decoherence functional 2}), and
writing 
$S_{\rm eff}[q_+,q_-]\equiv S_{\rm eff}
[q_{\scriptscriptstyle \Delta},q_{\scriptscriptstyle \Sigma}]$, we
get
\be
{\cal P}_{\bar{q}}( \{ \bar{q}_{\scriptscriptstyle \Sigma} \} )
_{(t_1, \dots , t_N)} \simeq
\int\! {\cal D}[q_{\scriptscriptstyle \Sigma}]\,
\prod_{k=1}^N 
\gamma^2 (q_{\scriptscriptstyle \Sigma}(t_k) \!-\! 
\bar{q}_{{\scriptscriptstyle \Sigma}_k}) \: 
{\cal P}_{\rm f}[q_{\scriptscriptstyle \Sigma}],
\label{probability 2}
\ee
where
\be
{\cal P}_{\rm f}[q_{\scriptscriptstyle \Sigma}] \equiv
\int_{_{_{\scriptstyle 
q_{\mbox{}_{\hspace{-0.1ex}\Delta}}\!(t_f)=0 }}}
\!\!\!\!\!\!\!\!\!\!
{\cal D}[q_{\scriptscriptstyle \Delta}]\; 
\rho_s \!\left(q_{\scriptscriptstyle \Sigma_{\scriptstyle i}}
\!+\!{\textstyle \frac{1}{2}}\, 
q_{\scriptscriptstyle \Delta_{\scriptstyle i}}, 
q_{\scriptscriptstyle \Sigma_{\scriptstyle i}}
\!-\!{\textstyle \frac{1}{2}}\, 
q_{\scriptscriptstyle \Delta_{\scriptstyle i}}
;t_i \right) \:
e^{{i \over \hbar}\, S_{\rm eff}
[q_{\mbox{}_{\hspace{-0.1ex}\Delta}},q_{\mbox{}_{\Sigma}}]}.
\label{probability 3}
\ee

At this stage, we introduce two simplifications in our analysis.
First, we restrict our evaluation to coarse-grained system variables
having significance only up to certain scales, larger enough 
than $\sigma$, so that the random variables $\bar{q}_k$ can be well
approximated by continuous random variables. This approximation can be
implemented with the use of a set of approximate projection operators
$\hat{P}_{\bar{q}}(t)$, with $\bar{q}$ being continuous variables,
which satisfy the properties (\ref{proj properties}) in an approximate
way (see Refs.~\cite{dowker,halliwell} for an example).
Then, all the sums $\sum_{\{ \bar{q}\} }$ can be replaced by integrals
$\int \hspace{-0.2ex} \prod_{k=1}^N d\bar{q}_k$ and 
the functions
${\cal P}_{\bar{q}}( \{ \bar{q} \} )_{(t_1, \dots , t_N)}$
become probability densities.
Second, as long as we are only interested in the dynamics of the
system on time 
scales much larger than $\Delta t_c$ ($\Delta t_c$ is proportional to
the decoherence time scale $t_{\scriptscriptstyle \! D}$, which is
typically extremely small, see Refs.~\cite{omnes,zurek,joos-zeh} for
some examples), we can take the continuous time limit in 
(\ref{probability 2}). 
In order to do so, consider the instants 
$t_k \equiv t_i+k \, \Delta t$, $k = 0, \dots , N+1$, with
$\Delta t \equiv (t_f-t_i)/(N+1)$.
Introducing functions $\bar{q}(t)$, such that $\bar{q}(t_k) =
\bar{q}_k$
(assumed now to be continuous variables), 
and letting $N \!\rightarrow \! \infty$ in 
(\ref{probability 2}) [replace 
$\bar{q}_{{\scriptscriptstyle \Sigma}_k}$ by $\bar{q}_k$],
with $(t_f-t_i)$ maintained finite (thus,  
$\Delta t \!\rightarrow \! 0$), we get a probability distribution
functional associated to some stochastic variables 
$\bar{q}(t)$ \cite{feynman-hibbs}:
\be
{\cal P}_{\bar{q}}[\bar{q}] \simeq
\int\! {\cal D}[q_{\scriptscriptstyle \Sigma}]\:
\gamma^2[q_{\scriptscriptstyle \Sigma}
-\bar{q}] \: 
{\cal P}_{\rm f}[q_{\scriptscriptstyle \Sigma}],
\label{prob functional}
\ee
where $\gamma[q]$ is the functional corresponding to 
$\prod_{k=1}^N \gamma (q(t_k))$ in the limit 
$N \!\rightarrow \!\infty$ (some redefinitions in the parameters
entering in the function $\gamma (q)$ may be needed in order that such
limit is well defined, see Refs.~\cite{halliwell,halliwell93} 
for an explicit example of how this limit is taken).
Notice that, if we take the limit to the continuous in time and
in the variables $\bar{q}_k$ in 
(\ref{decoherence condition}), we get a functional $\Phi_{\bar{q}}[j]$
which is the functional Fourier transform of 
${\cal P}_{\bar{q}}[\bar{q}]$. Hence, $\Phi_{\bar{q}}[j]$ 
can be interpreted as the 
characteristic functional for the stochastic variables 
$\bar{q}(t)$ \cite{feynman-hibbs}. 
From the probability functional (\ref{prob functional}) or,
equivalently, from the associated characteristic functional [by
functional derivation with respect to the sources $j(t)$],
we can compute the Green functions
$G_{c \; m_1 \cdots \, m_s}^{n_1 \cdots \, n_r} 
(t_1^{\prime}, \dots, t_r^{\prime};
t_1^{\prime \prime}, \dots, t_s^{\prime \prime})$ with each of the 
instants in $\{t_1^{\prime}, \dots , t_r^{\prime} \}$ 
being separated 
from $t_i$ and from the remaining instants in this set by
intervals larger enough than $\Delta t_c$, and similarly for 
the instants in 
$\{t_1^{\prime \prime}, \dots , t_s^{\prime \prime} \}$.

We can get a good approximation to the path integral 
(\ref{probability 3}) by expanding 
$S_{\rm eff}
[q_{\scriptscriptstyle \Delta},q_{\scriptscriptstyle \Sigma}]$ in
powers of $q_{\scriptscriptstyle \Delta}$ and neglecting higher than
quadratic terms, {\it i.e.}, we make a Gaussian approximation in this
path integral. This expansion can be made using 
(\ref{effective action 2}) and writing
$S_{\rm IF}^{\rm eff}[q_+,q_-]\equiv 
S_{\rm IF}^{\rm eff}
[q_{\scriptscriptstyle \Delta},q_{\scriptscriptstyle \Sigma}]$.
In this expansion, the dependence of
$S_{\rm eff}
[q_{\scriptscriptstyle \Delta},q_{\scriptscriptstyle \Sigma}]$ 
on the velocities 
$\dot{q}_{\scriptscriptstyle \Delta}(t)$\footnote{We understand that a
term depends on
$\dot{q}_{\scriptscriptstyle \Delta}(t)$ if it does so before 
any integration by parts.}
(we assume that there is no dependence on time derivatives of higher
order) gives rise, after integration by parts, to boundary terms 
proportional to $q_{{\scriptscriptstyle \Delta}_{\scriptstyle i}}$
(we use that $q_{{\scriptscriptstyle \Delta}_{\scriptstyle f}}=0$). 
For instance, assuming that 
$S_{s}^{\rm eff}[q]= \int\! dt \, L_s(q(t),\dot{q}(t),t)$, in the
expansion of the terms $S_{s}^{\rm eff}$ we find a boundary term 
$- p_{s}(q_{\scriptscriptstyle \Sigma_{\scriptstyle i}},
\dot{q}_{\scriptscriptstyle \Sigma_{\scriptstyle i}},t_i) \,
q_{\scriptscriptstyle \Delta_{\scriptstyle i}}$,
where $p_{s} \equiv \partial L_s/ \partial \dot{q}$ are the canonical
momenta. Similarly, if $S_{\rm IF}^{\rm eff}$ depends on
$\dot{q}_{\scriptscriptstyle \Delta}(t)$, its expansion will contain
some boundary terms. 
However, since, in general, 
$S_{\rm IF}^{\rm eff}$ depends non-locally on 
$q_{\scriptscriptstyle \Delta}(t)$ and 
$q_{\scriptscriptstyle \Sigma}(t)$, these terms will be more
complicated. 
Note that we are considering models slightly more general than the
ones studied by Gell-Mann and Hartle in 
Refs.~\cite{gell-mann-hartle,hartle}, since we allow for the
possibility of an influence action depending on 
$\dot{q}_{\scriptscriptstyle \Delta}(t)$ and
$\dot{q}_{\scriptscriptstyle \Sigma}(t)$. 
The motivation for considering such a generalization is that we are
interested in field theory actions with interaction terms depending
on the derivatives of the fields.

One can show that, when expanding up to quadratic order
in $q_{\scriptscriptstyle \Delta}$, the general form for the boundary
terms in $S_{\rm IF}^{\rm eff}$ is
$- F_1[q_{\scriptscriptstyle \Sigma}](t_i) \, 
q_{\scriptscriptstyle \Delta_{\scriptstyle i}}+
i F_2[q_{\scriptscriptstyle \Sigma}](t_i) \, 
q_{\scriptscriptstyle \Delta_{\scriptstyle i}}^2
+i \!\int\! dt \, q_{\scriptscriptstyle \Delta}(t) \, 
F_3[q_{\scriptscriptstyle \Sigma}](t,t_i) \,
q_{\scriptscriptstyle \Delta_{\scriptstyle i}}$,
where $F_1$, $F_2$ and $F_3$ are real functionals of 
$q_{\scriptscriptstyle \Sigma}$, which vanish when 
$S_{\rm IF}^{\rm eff}$ does not depend on 
$\dot{q}_{\scriptscriptstyle \Delta}(t)$. 
Finally, we get the following expansion:
\bea
&& \hspace{-3.4ex}
S_{\rm eff}
[q_{\scriptscriptstyle \Delta},q_{\scriptscriptstyle \Sigma}]=
S_s^{\rm eff}[q_{\scriptscriptstyle \Sigma}
\!+\!{\textstyle \frac{1}{2}}\, q_{\scriptscriptstyle \Delta}]
-S_s^{\rm eff}[q_{\scriptscriptstyle \Sigma}
\!-\!{\textstyle \frac{1}{2}}\, q_{\scriptscriptstyle \Delta}]
+S_{\rm IF}^{\rm eff}
[q_{\scriptscriptstyle \Delta},q_{\scriptscriptstyle \Sigma}]
= -p_1[q_{\scriptscriptstyle \Sigma}](t_i) \, 
q_{\scriptscriptstyle \Delta_{\scriptstyle i}}+
i F_2[q_{\scriptscriptstyle \Sigma}](t_i) \, 
q_{\scriptscriptstyle \Delta_{\scriptstyle i}}^2
    \nn  \\
&&\hspace{-3.4ex}
+\, i \!\int\! dt \, q_{\scriptscriptstyle \Delta}(t) \, 
F_3[q_{\scriptscriptstyle \Sigma}](t,t_i) \,
q_{\scriptscriptstyle \Delta_{\scriptstyle i}}
\!+\!\! \int\! dt\:  q_{\scriptscriptstyle \Delta}(t)
C[q_{\scriptscriptstyle \Sigma}](t) 
+{i \over 2\hbar} \!\int\! dt\, dt^{\prime}\,
q_{\scriptscriptstyle \Delta}(t) \, 
q_{\scriptscriptstyle \Delta}(t^{\prime}) \, 
C_2[q_{\scriptscriptstyle \Sigma}](t,t^{\prime})
+O \!\left(q_{\scriptscriptstyle \Delta}^3 \right) \!, 
 \nn  \\
\mbox{}
\label{eff action expansion}
\eea
with
\be
p_1[q](t_i) \equiv p_s (q_i,\dot{q}_i,t_i)+F_1[q](t_i),
\hspace{5ex}
C[q](t) \equiv {\delta S_s^{\rm eff}[q] 
\over \delta q(t)}+C_1[q](t),
\label{C}
\ee
and
\be
C_k[q_{\scriptscriptstyle \Sigma}](t_1,.\,.\,.\,,t_k)\equiv
\left( {\hbar \over i} \right)^{\! k-1} \!\!\!
\left. {\delta^k S_{\rm IF}^{\rm eff}
[q_{\scriptscriptstyle \Delta},q_{\scriptscriptstyle \Sigma}]
\over \delta q_{\scriptscriptstyle \Delta}(t_1)
\cdot \cdot \cdot \delta q_{\scriptscriptstyle \Delta}(t_k)}
\right|_{q_{\mbox{}_{\hspace{-0.1ex}\Delta}}=0},
\label{C's}
\ee
where the functional derivatives with respect to $q(t)$ are defined
for variations which keep the value of $q(t)$ fixed at $t=t_i$
and  $t=t_f$.

Substituting the expansion (\ref{eff action expansion})
into Eq.~(\ref{probability 3}), we 
get a Gaussian path integral, which can be calculated. 
Note that, since ${\rm Im}\, S_{\rm IF}^{\rm eff} \geq 0$,
$C_2[q](t,t^{\prime})$ is positive semi-definite. In order that 
the Gaussian approximation that we have carried out is valid, we must  
assume in addition that $C_2[q](t,t^{\prime})$ is strictly positive
definite and, thus, $\det C_2[q] \neq 0$.
We get
\be
{\cal P}_{\rm f}[q] \simeq 
N \,
W_i[q]  \left[ \det \!
\left( C_2[q]/ 2 \pi \hbar^2 \right) \right]^{-1/2} 
e^{-{1 \over 2}\! \int\! dt\, dt^{\prime}\, C[q](t)\,
C_2^{-1}[q](t,t^{\prime})\, C[q](t^{\prime}) },
\label{approx probability}
\ee
where $N$ is a normalization constant, 
$C_2^{-1}$ is the inverse of
$C_2$ defined by 
\be
\int \! dt^{\prime \prime}\, C_2(t,t^{\prime \prime}) \,
C_2^{-1}(t^{\prime \prime},t^{\prime})= \delta(t-t^{\prime}),
\ee
$W_i[q] \equiv W\! \left(q(t_i),p[q](t_i),\Pi[q](t_i);t_i \right)$, 
with 
\be
W(q,p,\Pi;t_i) 
\equiv \int\! {dq_0 \over 2 \pi \hbar } \;
e^{-{i \over \hbar} q_0 p} \,
e^{-{1 \over \hbar} q_0^2 \Pi}
\rho_s (q_{\scriptstyle i}+{\textstyle \frac{1}{2}}
\hspace{0.2ex} q_0, 
q_{\scriptstyle i}-{\textstyle \frac{1}{2}}
\hspace{0.2ex} q_0;t_i ),
\label{Wigner funct} 
\ee 
and 
\bea
&&p[q](t_i) \equiv p_1[q](t_i)+ \hbar \! \int \! dt \, dt^{\prime } \,
F_3[q](t,t_i) \, C_2^{-1}[q](t,t^{\prime})\, C[q](t^{\prime}),
\nn  \\
&&\Pi[q](t_i) \equiv F_2[q](t_i)
- {\hbar \over 2}\! \int \! dt \, dt^{\prime } \,
F_3[q](t,t_i) \, C_2^{-1}[q](t,t^{\prime})\, F_3[q](t^{\prime},t_i). 
\eea
Note that the function $W$ defined in (\ref{Wigner funct}) is a
generalization of the Wigner function associated to the initial state
of the system, and it reduces to the ordinary Wigner function for
$\Pi =0$ \cite{wigner}.
Note that, in expression (\ref{approx probability}),
the momenta $p[q](t_i)$ in this generalized Wigner function are in
general different from the canonical momenta $p_s (q_i,\dot{q}_i,t_i)$. 
In the case of $S^{\rm eff}_{\rm IF}$
non-depending on the velocities 
$\dot{q}_{\scriptscriptstyle \Delta}(t)$, one has
$p[q](t_i)=p_s (q_i,\dot{q}_i,t_i)$ and $\Pi[q](t_i)=0$, thus,
$W_i[q]$ is the standard Wigner function. 
From the definition (\ref{C's}), and using the properties of
$S_{\rm IF}^{\rm eff}[q_+,q_-]$, we can see that
\bea
C_1[q](t)\!\!\!&=&\!\!\!
\left.{\delta \, {\rm Re}\, S_{\rm IF}^{\rm eff}[q_+,q_-]
\over \delta q_+(t) }\right|_{q_+=q_-=q}
=\left.{\delta S_{\rm IF}^{\rm eff}[q_+,q_-]
\over \delta q_+(t) }\right|_{q_+=q_-=q},  \nn  \\
C_2[q](t,t^{\prime})\!\!\!&=&\!\!\!{\hbar \over 2} \left.
\left[ {\delta^2\, {\rm Im}\, S_{\rm IF}^{\rm eff}[q_+,q_-]
\over \delta q_+(t) \delta q_+(t^{\prime}) }
-{\delta^2\, {\rm Im}\, S_{\rm IF}^{\rm eff}[q_+,q_-]
\over \delta q_+(t) \delta q_-(t^{\prime}) } \right] 
\right|_{q_+=q_-=q},
\label{C's 2}
\eea
and then, from (\ref{C}) and (\ref{effective action 2}), we have
\be
C[q](t)=\left.{\delta S_{\rm eff}[q_+,q_-]
\over \delta q_+(t) }\right|_{q_+=q_-=q}.
\ee

Substituting (\ref{approx probability}) into
(\ref{prob functional}), we see that the only non-negligible 
contribution to the path integral
in (\ref{prob functional}) come from those paths which are
not very far deviated from the paths 
$q^{\scriptscriptstyle (0)}(t)$ which
satisfy $C[q^{\scriptscriptstyle (0)}](t)=0$, that is, which satisfy
the semiclassical equation (\ref{semiclassical eq}).
This implies that only those paths $\bar{q}(t)$ which remain
always near from the semiclassical paths 
$q^{\scriptscriptstyle (0)}(t)$ will give a non-negligible value to
${\cal P}_{\bar{q}}[\bar{q}]$.
In this sense, the mechanism proposed by Gell-Mann and Hartle
is a mechanism for decoherence and classicalization of coarse-grained
system variables. 
However, we see that, in general, ${\cal P}_{\bar{q}}[\bar{q}]$
has a complicated functional dependence on $\bar{q}(t)$.

Let us then study the deviations from a specific solution of the 
semiclassical equation, that is, we shall now restrict our
considerations to those paths $\bar{q}(t)$ which are distributed
around a given solution $q^{\scriptscriptstyle (0)}(t)$ of the
semiclassical equation. 
We can now introduce stochastic variables 
$\Delta q(t) \equiv \bar{q}(t)- q^{\scriptscriptstyle (0)}(t)$
which describe
the deviations from $q^{\scriptscriptstyle (0)}(t)$.
The associated probability distribution functional 
${\cal P}_{\!\Delta q}[\Delta q]$ is equal to 
${\cal P}_{\bar{q}}[q^{\scriptscriptstyle (0)}\!+\!\Delta q]$
up to a normalization factor, which, 
from (\ref{prob functional}), is given by 
\be
{\cal P}_{\bar{q}}[q^{\scriptscriptstyle (0)}+\Delta q]
\simeq
\int\! {\cal D}[q]\:
\gamma^2[q] \: 
{\cal P}_{\rm f}[q^{\scriptscriptstyle (0)} +\Delta q+q].
\label{probability 4}
\ee

In practice, it is difficult 
to work out the explicit dependence of the 
probability distribution functional on the 
characteristic parameters of the coarse-graining, 
$\sigma$ and $\Delta t_c$, even in simple models 
\cite{halliwell93,halliwell}. 
Nevertheless, if such parameters are small enough so that the values
of ${\cal P}_{\rm f}[q^{\scriptscriptstyle (0)} +\Delta q+q]$
do not change very much for the different paths 
$q(t)$ which give a non-negligible contribution in
(\ref{probability 4}),  
the functional (\ref{probability 4}) can be approximated by
${\cal P}_{\rm f}[q^{\scriptscriptstyle (0)} +\Delta q]$. 
We can make a further approximation by expanding 
${\cal P}_{\rm f}[q^{\scriptscriptstyle (0)}+\Delta q]$ 
around $q^{\scriptscriptstyle (0)}$. 
This can be done by setting  
$q_{\scriptscriptstyle \Sigma}= q^{\scriptscriptstyle (0)}
+\Delta q$ in (\ref{eff action expansion}), expanding in $\Delta q$,
and substituting the result for this expansion in
(\ref{probability 3}).
The result to lowest non-trivial order is 
\be
{\cal P}_{\!\Delta q}[\Delta q] 
\simeq N[q^{\scriptscriptstyle (0)}] \, 
W_i[q^{\scriptscriptstyle (0)}\!+\!
\Delta q] \,
e^{-{1 \over 2} \int\! dt\, dt^{\prime}\, 
C_L[q^{\scriptscriptstyle (0)}+\Delta q](t)\,
C_2^{-1}[q^{\scriptscriptstyle (0)}](t,t^{\prime})\, 
C_L[q^{\scriptscriptstyle (0)}+\Delta q](t^{\prime}) },
\label{approx gaussian probability}
\ee
where $N[q^{\scriptscriptstyle (0)}]$ is a normalization factor and
$C_L[q^{\scriptscriptstyle (0)}+\Delta q]$ is
the expansion of $C[q^{\scriptscriptstyle (0)}+\Delta q]$
to linear order in $\Delta q$.
Notice that, in this probability functional, 
the factor 
$W_i[q^{\scriptscriptstyle (0)}\!+\!
\Delta q]$ contains all the contribution arising from the
initial state of the system. This generalized Wigner function, even if
computed expanding around $q^{\scriptscriptstyle (0)}$, will have
in general a complicated non-local dependence on $\Delta q$, except
when $S^{\rm eff}_{\rm IF}$ is independent of   
$\dot{q}_{\scriptscriptstyle \Delta}$, in which case it reduces to the
standard Wigner function for the initial state of the system and
depends only on $\Delta q_i$ and $\Delta \dot{q}_i$.
If the deviations from $q^{\scriptscriptstyle (0)}$ are small enough,
we can approximate $W_i[q^{\scriptscriptstyle (0)}+
\Delta q] \simeq W_i[q^{\scriptscriptstyle (0)}]$.
Then, with these approximations, the variables $\Delta q$ are
distributed in such a way that 
$C_L[q^{\scriptscriptstyle (0)}\!+\!\Delta q](t)$ are
Gaussian stochastic variables characterized by
\be
\left\langle C_L[q^{\scriptscriptstyle (0)}\!+\!\Delta q](t)
\right\rangle_c=0,
\hspace{7ex}
\left\langle C_L[q^{\scriptscriptstyle (0)}\!+\!\Delta q](t)\,
C_L[q^{\scriptscriptstyle (0)}\!+\!\Delta q](t^{\prime})
\right\rangle_c=
C_2[q^{\scriptscriptstyle (0)}](t,t^{\prime}) .
\label{gaussian correlators}
\ee
Thus, the equation of motion for $\Delta q$ is the Langevin equation
\be
C_L[q^{\scriptscriptstyle (0)}\!+\!\Delta q](t)
+\xi(t)=0,
\label{langevin eq}
\ee
where $\xi(t)$ is a Gaussian stochastic source with
\be
\left\langle \xi(t)
\right\rangle_c=0,
\hspace{7ex}
\left\langle \xi(t) \,
\xi(t^{\prime})
\right\rangle_c= 
C_2[q^{\scriptscriptstyle (0)}](t,t^{\prime}) .
\label{gaussian correlators 2}
\ee

We should mention that there are very simple models for quantum
Brownian motion in which all the actions involved are quadratic
in their variables and the interaction terms are independent of the
velocities 
\cite{feynman-vernon,feynman-hibbs,caldeira,hu-paz-zhang,%
gell-mann-hartle,hartle,halliwell93,dowker,grabert,brun}.
For such models, assuming that the environment is in an initial state
of thermal equilibrium, the influence functional can be computed
exactly and it is Gaussian. The effective action of Feynman
and Vernon in these
cases is exactly of the form (\ref{eff action expansion}), with
$C_1[q_{\scriptscriptstyle \Sigma}](t)$ linear in 
$q_{\scriptscriptstyle \Sigma}$, $C_2(t,t^{\prime})$ independent
of $q_{\scriptscriptstyle \Sigma}$ and $F_1\!=\!F_2\!=\!F_3\!=\!0$. 
Thus, for these models,
expression (\ref{approx probability}) is actually exact. 
In these cases, with the approximation
${\cal P}_{\bar{q}}[\bar{q}] \simeq {\cal P}_{\rm f}[\bar{q}]$,
one can derive a Langevin equation for the stochastic variables
$\bar{q}(t)$, without need of introducing a specific solution
$q^{\scriptscriptstyle (0)}$ of the semiclassical equation. 
This Langevin
equation is simply $C[\bar{q}](t)\!+\!\xi(t)\!=\!0$, being $\xi(t)$ a
Gaussian stochastic source with $\left\langle \xi(t)\right\rangle_c=0$
and $\left\langle \xi(t) \, \xi(t^{\prime})\right\rangle_c= 
C_2(t,t^{\prime})$.  
However, for models with more complicated actions, 
we are only able to derive effective equations of motion 
for the deviations $\Delta q$ around a given solution 
$q^{\scriptscriptstyle (0)}$ of the semiclassical equation.


\subsection{\hspace{-2.5ex}. A quick method to obtain 
the Langevin equation}
\label{subsec:quick method}


Starting with the effective action of Feynman and Vernon 
(\ref{effective action 2}), there
is a quick way to obtain the Langevin equation (\ref{langevin eq})
for the deviations $\Delta q$ around a specific solution of 
the semiclassical equation. 
This method has actually been extensively used in the
literature, in the context of quantum Brownian motion 
\cite{caldeira,hu-matacz94,hu-paz-zhang2}, 
and also in the context of field theory 
\cite{greiner,matacz,morikawa,shaisultanov,gleiser}, 
including some models for gravity interacting with a scalar field
\cite{calzettahu,humatacz,husinha,cv96,lomb-mazz,ccv97,campos-hu}.
One starts with an expansion of this effective action around a
solution $q^{\scriptscriptstyle (0)}(t)$ of the semiclassical equation
up to quadratic order in perturbations $\Delta q_{\pm}$
satisfying $\Delta q_+(t_i)=\Delta q_-(t_i)$ and 
$\Delta q_+(t_f)=\Delta q_-(t_f)$ 
(in the simplest models, in
which this effective action is exactly quadratic in $q_+$ and $q_-$,
one works directly with the exact expression). 
From (\ref{eff action expansion}), it is easy to see that
the expansion for the influence action reads
\bea
&&\hspace{-10.5ex}
S^{\rm eff}_{\rm IF}[q^{\scriptscriptstyle (0)}\!+\!\Delta q_+ ,
q^{\scriptscriptstyle (0)}\!+\!\Delta q_-]=
\int\! dt  \left(\Delta q_+(t)\!-\!\Delta q_-(t) \right)
C_1[q^{\scriptscriptstyle (0)}\!+\!{\textstyle \frac{1}{2}}
\hspace{0.2ex}(\Delta q_+ \!+\! \Delta q_-)](t)  \nn  \\
&& \hspace{-1ex}
+\,{i \over 2\hbar} \int\! dt\, dt^{\prime}
\left(\Delta q_+(t)\!-\!\Delta q_-(t) \right)
C_2[q^{\scriptscriptstyle (0)}](t,t^{\prime})
\left(\Delta q_+(t^{\prime})\!-\!\Delta q_-(t^{\prime}) \right)
+O (\Delta q^3 ), 
\label{influence action expansion}
\eea
where it is understood that $C_1$ has to be
expanded up to linear order. Using the identity, which follows from a
Gaussian path integration, 
\be
e^{-{1  \over 2\hbar^2}\! \int\! dt\, dt^{\prime}\,
\left(\Delta q_+(t)-\Delta q_-(t) \right)\,
C_2[q^{\scriptscriptstyle (0)}](t,t^{\prime})\,
\left(\Delta q_+(t^{\prime})-\Delta q_-(t^{\prime}) \right)}=
\int\! {\cal D}[\xi]\: {\cal P}_{\xi}[\xi]\, e^{{i \over \hbar} 
\!\int\! dt\, \xi(t)\, 
\left(\Delta q_+(t)-\Delta q_-(t) \right) },
\label{identity}
\ee
where ${\cal P}_{\xi}[\xi]$ is the Gaussian probability distribution
functional for the Gaussian stochastic variables
$\xi(t)$ characterized by
(\ref{gaussian correlators 2}), that is,
\be
{\cal P}_{\xi}[\xi]=
 \frac{e^{-{1\over2}\!\int\! dt\, dt^{\prime}\, 
 \xi(t) \,
 C_2^{-1}[q^{\scriptscriptstyle (0)}](t,t^{\prime})\,
 \xi(t^{\prime}) }}
 {\int\! {\cal D}\bigl[\bar{\xi}\bigr]\:
   e^{-{1\over2}\!\int\! d\tau \, d\tau ^{\prime}\,
\bar{\xi}(\tau) \,
 C_2^{-1}[q^{\scriptscriptstyle (0)}](\tau,\tau ^{\prime})\,
\bar{\xi}(\tau ^{\prime}) }},
\label{xi probability}
\ee
we can write in this approximation
\be
\bigl|\hspace{0.2ex}{\cal F}^{\rm eff}_{\rm IF}
[q^{\scriptscriptstyle (0)}\!+\!\Delta q_+,
q^{\scriptscriptstyle (0)}\!+\!\Delta q_-]
\hspace{0.2ex}\bigr|=
e^{-{1 \over \hbar}\,{\rm Im}\, S^{\rm eff}_{\rm IF}
[q^{\scriptscriptstyle (0)}+\Delta q_+,
q^{\scriptscriptstyle (0)}+\Delta q_-]}=
\left\langle e^{{i \over \hbar}\! \int\! dt \,
\xi(t)\, \left(\Delta q_+(t)-\Delta q_-(t) \right) }
\right\rangle_c,
\label{identity2}
\ee
where $\langle \hspace{1.5ex} \rangle_c$ means statistical average
over the stochastic variables $\xi(t)$.
Thus, the effect of the imaginary part of the influence action 
(\ref{influence action expansion}) on the corresponding influence
functional is equivalent to the averaged effect of the stochastic
source $\xi(t)$ coupled linearly 
to the perturbations $\Delta q_{\pm}$
(note that, in the above expressions, the perturbations
$\Delta q_{\pm}$ are deterministic functions).
Notice that expression (\ref{identity}) or, equivalently, 
(\ref{identity2}) give the characteristic functional of the
stochastic variables $\xi(t)$ \cite{feynman-hibbs}.
The influence functional, in the approximation
(\ref{influence action expansion}), can then be written as an
statistical average over $\xi$:
\be
{\cal F}^{\rm eff}_{\rm IF}
[q^{\scriptscriptstyle (0)}\!+\!\Delta q_+,
q^{\scriptscriptstyle (0)}\!+\!\Delta q_-]=
\left\langle e^{{i \over \hbar}\, 
{\cal A}^{\rm eff}_{\rm IF}[\Delta q_+,\Delta q_-;\xi] }
\right\rangle_c,
\ee
where 
\be
{\cal A}^{\rm eff}_{\rm IF}[\Delta q_+,\Delta q_-;\xi] \equiv 
{\rm Re}\, S^{\rm eff}_{\rm IF}
[q^{\scriptscriptstyle (0)}\!+\!\Delta q_+,
q^{\scriptscriptstyle (0)}\!+\!\Delta q_-]
+\!\int\! dt \,
\xi(t)
\left(\Delta q_+(t)-\Delta q_-(t) 
\right) +O (\Delta q^3 ),
\ee
where ${\rm Re}\, S^{\rm eff}_{\rm IF}$ can be read from
expression (\ref{influence action expansion}). The Langevin equation
(\ref{langevin eq}) can be easily derived from the action
\be
{\cal A}_{\rm eff}[\Delta q_+,\Delta q_-;\xi] \equiv 
S_s^{\rm eff}[q^{\scriptscriptstyle (0)}\!+\!\Delta q_+]-
S_s^{\rm eff}[q^{\scriptscriptstyle (0)}\!+\!\Delta q_-]+
{\cal A}^{\rm eff}_{\rm IF}[\Delta q_+,\Delta q_-;\xi],
\label{new effective action}
\ee
where $S_s^{\rm eff}[q^{\scriptscriptstyle (0)}\!+\!\Delta q_{\pm}]$
has to be expanded up to second order in the perturbations 
$\Delta q_{\pm}$. That is, 
\be
\left. \frac{\delta {\cal A}_{\rm eff}[\Delta q_+,\Delta q_-;\xi]}
{\delta \Delta q_+(t)}
\right|_{\Delta q_+=\Delta q_-=\Delta q}=0
\label{quick Langevin eq}
\ee
leads to Eq.~(\ref{langevin eq}).


\section{\hspace{-2.5ex}. Effective equations of motion 
for the gravitational field}
\label{sec:Einstein-Langevin}


\setcounter{equation}{0}

In this section, we shall apply the results of the previous section to
derive effective equations of motion for the gravitational field in a
semiclassical regime. In order to do so, we will consider 
the simplest case of a linear real scalar field $\Phi$ coupled to
the gravitational field.
We shall restrict ourselves to the case of fields defined on a
globally hyperbolic manifold ${\cal M}$.
In this case, we would consider
the metric field $g_{ab}(x)$ as the system degrees of freedom, 
and the scalar field $\Phi(x)$ 
and also some ``high-momentum'' gravitational modes, 
considered as inaccessible to the observations, as the
environment variables. 
Unfortunately, since the form of a complete
quantum theory of gravity interacting with matter is unknown,
we do not know what these ``high-momentum'' gravitational modes are.
Such a fundamental quantum theory might not even be a field theory,
in which case the metric and scalar fields would not be 
fundamental objects. 
Thus, in this case, we cannot attempt to evaluate 
the effective actions in Eq.~(\ref{s-e effective actions}) starting
from the fundamental quantum theory and integrating out the
``high-momentum'' gravitational modes .
What we can do instead is to adopt the usual procedure when dealing
with an effective quantum field theory. That is, we shall take for 
the actions $S_{s}^{\rm eff}[g]$ and 
$S_{se}^{\rm eff}[g^+,\Phi_+;g^-,\Phi_-]$ the most general local form
compatible with general covariance and with the properties of
$S_{se}^{\rm eff}$ [these properties are analogous to those of 
$S_{\rm IF}$ in Eq.~(\ref{influence action properties})] 
\cite{weinberg,donoghue}.
The general form for $S_{s}^{\rm eff}[g]$ is 
\be
S_{s}^{\rm eff}[g]=\int \! d^4 x \, \sqrt{- g} 
\left[{1\over 16 \pi G_{B}}
\left(R-2\Lambda_{B}\right)
+ \alpha_{B} C_{abcd}C^{abcd}+\beta_{B} R^2 + \cdots \right],
\label{grav action}
\ee
where $R$ and $C_{abcd}$ are, respectively, the scalar curvature and
the Weyl tensor associated to the metric $g_{ab}$, 
$1/G_B$, $\Lambda_B /G_B$, $\alpha_B$ and $\beta_B$ are
bare coupling constants and the dots represent
terms of higher order in the curvature [because of the
Gauss-Bonnet theorem in four spacetime dimensions, there is no need of
considering terms of second order in the curvature different from
those written in Eq.~(\ref{grav action})]. 
Since ${\cal M}$ is a globally hyperbolic manifold, 
we can foliate it by a family of Cauchy
hypersurfaces $\Sigma_{t}$, labeled by a time coordinate $t$. 
We use the notation 
${\bf x}$ for spatial coordinates on each of these hypersurfaces, and
$t_{i}$ and $t_{f}$ for some initial and final times, respectively.
The integration domain for all the action terms must now be understood
as a compact region ${\cal U}$ of the manifold ${\cal M}$, bounded by
the hypersurfaces $\Sigma_{t_i}$ and $\Sigma_{t_f}$
({\it i.e.}, as in the previous section, integrals in $t$ are  
integrals between $t_i$ and $t_f$).

For the matter part of the effective action, let us consider
the following ansatz:
\be
S_{se}^{\rm eff}[g^+,\Phi_+;g^-,\Phi_-]=S_m[g^+,\Phi_+]
-S_m[g^-,\Phi_-],
\label{effective action ansatz}
\ee
with
\be
S_m[g,\Phi] \equiv -{1\over2} \int\! d^4x \, \sqrt{- g} 
  \left[g^{ab}\partial_a \Phi \hspace{0.2ex} \partial_b \Phi
  +\left(m^2+ \xi R \right)\Phi^2 + \cdots \right], 
\label{scalar field action}
\ee
where $\xi$ is a dimensionless coupling parameter
of the field to the scalar curvature, and the dots  
stand for terms of higher order in the curvature and in the number of
derivatives of the scalar field.
Self-interaction terms for the scalar field could also be included
but, for simplicity, we shall ignore them in this paper.
One can see that general covariance and the properties of 
$S_{se}^{\rm eff}[g^+,\Phi_+;g^-,\Phi_-]$ imply that imaginary terms
and terms mixing the ``plus'' and ``minus'' fields in this action must
be necessarily non-local. Thus, within a local approximation, 
the ansatz (\ref{effective action ansatz}) is the most general form
for this action. We shall comment below some limitations of
this local approximation.

In order to simplify the analysis, we neglect the contributions of the
higher order terms not written in 
Eqs.~(\ref{grav action}) and (\ref{scalar field action}).
Assuming that the mass of the scalar field is much smaller
than the Planck mass, this is a good approximation in a regime
where all the characteristic curvature scales are far enough 
from the Planck scales. 
The terms in the gravitational Lagrangian density
proportional to $R^2$ and $C_{abcd}C^{abcd}$ need to be considered in
order to renormalize the matter one-loop ultraviolet divergencies.

Assuming the form (\ref{effective action ansatz}) for the matter part
of the effective action, we can now
introduce the corresponding effective influence functional 
as in Eq.~(\ref{effective influence functional}). 
Let us assume that the state of the scalar field 
in the Schr\"{o}dinger picture 
at the initial time $t\! =\! t_{i}$ is described by a density
operator $\hat{\rho}^{\rm \scriptscriptstyle S}(t_{i})$ 
(in the notation of the previous section, this was 
$\hat{\rho}^{\rm \scriptscriptstyle S}_e(t_{i})$, but, here, we
drop the subindex $e$ to simplify the notation).
If we now consider the theory of a scalar field quantized in a
classical background spacetime $({\cal M},g_{ab})$ through the action 
(\ref{scalar field action}), to this state it would correspond a
state in the Heisenberg picture described by a density operator
$\hat{\rho}[g]$. Let 
$\left\{ |\varphi(\mbox{\bf x})\rangle^{\rm \scriptscriptstyle S} 
\right\}$ 
be the basis of eigenstates of the Schr\"{o}dinger picture scalar
field operator $\hat{\Phi}^{\rm \scriptscriptstyle S}({\bf x})$:
$\hat{\Phi}^{\rm \scriptscriptstyle S}({\bf x})
\, |\varphi\rangle ^{\rm \scriptscriptstyle S}=
\varphi(\mbox{\bf x})
\, |\varphi\rangle^{\rm \scriptscriptstyle S}$.
The matrix elements of 
$\hat{\rho}^{\rm \scriptscriptstyle S}(t_{i})$ in this basis will be
written as 
$\rho_{i} \!\left[\varphi,\tilde{\varphi}\right] \equiv 
\mbox{}^{\rm \scriptscriptstyle S}
\langle \varphi|\,\hat{\rho}^{\rm \scriptscriptstyle S}(t_{i})
\, |\tilde{\varphi}\rangle^{\rm \scriptscriptstyle S}$. 
We can now introduce the effective influence functional as
\be
{\cal F}^{\rm eff}_{\rm IF}[g^+,g^-] \equiv
\int\! {\cal D}[\Phi_+]\;
{\cal D}[\Phi_-] \;
\rho_i \!\left[\Phi_+(t_i),\Phi_-(t_i) \right] \,
\delta\!\left[\Phi_+(t_f)\!-\!\Phi_-(t_f)  \right]\:
e^{i\left(S_{m}[g^+,\Phi_+]-S_{m}[g^-,\Phi_-]\right) },
\label{path integral}
\ee 
and the effective influence action will be given by
${\cal F}^{\rm eff}_{\rm IF}[g^+,g^-] \equiv
e^{i S^{\rm eff}_{\rm IF}[g^+,g^-]}$.

Of course, trying to show how the mechanism for decoherence
and classicalization of the previous section can work in this case
would involve some technical difficulties, such as introducing
diffeomorphism invariant coarse-grainings and eliminating properly the
gauge redundancy (with the use of some suitable Faddeev-Popov method)
in the path integrals. We are not going to deal with such issues in
this paper. We shall rather assume that they can be suitably
implemented without changing the main results for the effective
equations of motion.

Expression (\ref{path integral}) is actually formal, it is ill-defined
and must be regularized in order to get a meaningful quantity for
the influence functional. We shall formally assume that we can regularize
it using dimensional regularization, that is, that we can give sense
to Eq.~(\ref{path integral}) by
dimensional continuation of all the quantities that appear in this
expression. We should mention that, however,
when performing specific calculations, 
the dimensional regularization procedure may not be
in all cases the most suitable one. 
In this sense, one should understand the following derivation as being
formal. Using dimensional regularization, we must
substitute the action $S_m$ in (\ref{path integral}) 
by some generalization to $n$ spacetime
dimensions. This can be taken as 
\be
S_m[g,\Phi_{n}] = -{1\over2} \int\! d^n x \, \sqrt{- g} 
  \left[g^{ab} \partial_a \Phi_{n} \partial_b \Phi_{n}
  +\left(m^2+ \xi R \right)\Phi_{n}^2  \right],
\label{scalar action}
\ee  
where we use a notation in which we write a subindex $n$ in all
the quantities that have different physical dimensions than the 
corresponding physical quantities in the spacetime
of four dimensions. The quantities that do not carry a
subindex $n$ have the same physical dimensions than  the corresponding
ones in four spacetime  dimensions, although they should not be
confused with such physical quantities.
A quantity with a subindex $n$ can always be associated to another one
without a subindex $n$; these are related by some 
mass scale $\mu$, for instance, it is easy to see
that $\Phi_{n}=\mu^{{n-4\over 2}}\,\Phi$.

In order to write the effective equations for the
metric field in dimensional regularization,
we need to substitute the action (\ref{grav action}) by 
some suitable generalization to $n$ spacetime dimensions. We 
take
\be
S_{s}^{\rm eff}[g]=\mu^{n-4} \!\int \! d^n x \,\sqrt{- g} 
\left[{1\over 16 \pi G_{B}}
 \left(R-2\Lambda_{B}\right)+ {2\over 3}\,\alpha_{B} 
 \left(R_{abcd}R^{abcd}-
  R_{ab}R^{ab}  \right)+\beta_{B} R^2 \right],
\label{grav action in n}
\ee
where $R_{abcd}$ is the Riemann tensor and,
again, the mass parameter $\mu$ has been introduced in order to
get the correct physical dimensions. Using the Gauss-Bonnet theorem in
four spacetime dimensions, one can see that the action obtained by
setting $n\!=\!4$ in (\ref{grav action in n}) is equivalent to
(\ref{grav action}). 
The form of the action (\ref{grav action in n}) is suggested from the
Schwinger-DeWitt analysis of the divergencies in the stress-energy
tensor in dimensional regularization \cite{bunch}. 
The effective action of Feynman and Vernon 
(\ref{effective action 2}) is in our case given by
$S_{\rm eff}[g^+,g^-]= S_{s}^{\rm eff}[g^+]-S_{s}^{\rm eff}[g^-]
+S^{\rm eff}_{\rm IF}[g^+,g^-]$.
Since the action terms (\ref{scalar action}) and 
(\ref{grav action in n})
contain second order derivatives of the metric, one should also add
some boundary terms to them \cite{wald84,humatacz}. 
The effect of these
boundary terms is simply to cancel out the boundary terms that appear
when taking variations of $S_{\rm eff}[g^+,g^-]$ that keep the value
of $g^+_{ab}$ and $g^-_{ab}$ fixed on the boundary of ${\cal U}$. 
They guarantee that we can obtain an expansion for 
$S_{\rm eff}[g^+,g^-]$ analogous to (\ref{eff action expansion}), with
no extra boundary terms coming from the integration by parts of terms
containing second order derivatives of 
$g_{ab}^{\scriptscriptstyle \Delta} \equiv g^+_{ab}-g^-_{ab}$.
Alternatively, in order to obtain the effective equations for
the metric [equations analogous to (\ref{semiclassical eq}) and
(\ref{langevin eq})], we can work with the action terms 
(\ref{scalar action}) and (\ref{grav action in n}) (without boundary
terms) and neglect all boundary terms when taking variations with
respect to $g^{\pm}_{ab}$. From now on, all the functional derivatives
with respect to the metric must be understood in this sense.


\subsection{\hspace{-2.5ex}. The semiclassical 
Einstein equation}
\label{subsec:semiclassical Einstein}


From the action 
(\ref{scalar action}), we can define the stress-energy tensor
functional in the usual way 
\be
T^{ab}[g,\Phi_{n}](x) \equiv {2\over\sqrt{- g(x)}} \, 
   \frac{\delta S_m[g,\Phi_{n}]}{\delta g_{ab}(x)},
\label{s-t functional}
\ee
which yields
\be
T^{ab}[g,\Phi_n]=\bigtriangledown^{a}\Phi_n
\bigtriangledown^{b}\hspace{-0.1ex}
\Phi_n- {1\over 2}\, g^{ab} 
\bigtriangledown^{c} \hspace{-0.1ex}
\Phi_n \bigtriangledown_{\!c}\hspace{-0.1ex} \Phi_n 
-{1\over 2}\, g^{ab}\, m^2 \Phi_n^2 
+\xi \left( g^{ab} \Box
-\bigtriangledown^{a}\! \bigtriangledown^{b}
+\, G^{ab} \right) \Phi_n^2
\label{class s-t} 
\ee
where $\bigtriangledown_{\!a}$ is the covariant derivative associated
to the metric $g_{ab}$,
$\Box \!\equiv\! \bigtriangledown_{\!a}\bigtriangledown^{a}$,
and $G_{ab}$ is the Einstein tensor. 
Working in the Heisenberg picture, we can now formally introduce the
stress-energy tensor operator for a scalar field quantized in a
classical spacetime background, regularized using dimensional
regularization, as 
\be
\hat{T}_{n}^{ab}[g] \equiv  T^{ab}[ g,\hat{\Phi}_{n}[g]], 
\hspace{5 ex}  
\hat{T}^{ab}[g] \equiv  \mu^{-(n-4)}\, \hat{T}_{n}^{ab}[g],
\label{regul s-t}
\ee
where $\hat{\Phi}_{n}[g](x)$ is the Heisenberg picture field operator
in $n$ spacetime dimensions, which satisfies the Klein-Gordon equation
\be
\left( \Box -m^2- \xi R \right) \hat{\Phi}_{n}=0,
\label{Klein-Gordon in n}
\ee
and where we use a symmetrical ordering (Weyl ordering) prescription
for the operators. Using 
Eq.~(\ref{Klein-Gordon in n}), one can write the stress-energy
operator in the following way:
\be
\hat{T}_{n}^{ab}[g] = {1\over 2} \left\{
     \bigtriangledown^{a}\hat{\Phi}_{n}[g]\, , \,
     \bigtriangledown^{b}\hat{\Phi}_{n}[g] \right\}
     + {\cal D}^{ab}[g]\, \hat{\Phi}_{n}^2[g],
\label{regul s-t 2}
\ee
where ${\cal D}^{ab}[g]$ is the differential operator
\be
{\cal D}^{ab}_{x} \equiv \left(\xi-{1\over 4}\right) g^{ab}(x) 
\Box_{x}+ \xi
\left( R^{ab}(x)- \bigtriangledown^{a}_{x} 
\bigtriangledown^{b}_{x} \right),
\label{diff operator}
\ee
being $R_{ab}$ the Ricci tensor.
From the definitions (\ref{path integral}),
(\ref{s-t functional}) and (\ref{regul s-t}), one
can see that 
\be
\left. {2\over\sqrt{- g(x)}} \, 
 \frac{\delta S^{\rm eff}_{\rm IF}[g^+,g^-]}
{\delta g^+_{ab}(x)} \right|_{g^+=g^-=g} \!
=\left\langle \hat{T}_n^{ab}(x) \right\rangle \![g],
\label{s-t expect value}
\ee 
where the expectation value is taken in the $n$-dimensional spacetime
generalization of the state described by
$\hat{\rho}[g]$.

As in Eq.~(\ref{semiclassical eq}), if we derive 
$S_{\rm eff}[g^+,g^-]$ with respect to $g^+_{ab}$ and then set
$g^+_{ab}=g^-_{ab}=g_{ab}$, we get
the semiclassical Einstein
equation in dimensional regularization:
\be
{1\over 8 \pi G_{B}} \left( G^{ab}[g]+ \Lambda_{B} g^{ab} \right)-
\left({4\over 3}\, \alpha_{B} D^{ab} 
+2  \beta_{B} B^{ab}\right)\! [g]
= \mu^{-(n-4)}
\left\langle \hat{T}_{n}^{ab}\right\rangle \! [g], 
\label{semiclassical eq in n} 
\ee
where the tensors $D^{ab}$ and $B^{ab}$ are defined as 
\begin{eqnarray}
&&\hspace{-6ex}
D^{ab} \equiv
{1\over\sqrt{- g}}   \frac{\delta}{\delta g_{ab}}
          \int \! d^n x \,\sqrt{- g} \left(R_{cdef}R^{cdef}-
                                       R_{cd}R^{cd}  \right)
    = {1\over2}\, g^{ab} \! \left(  R_{cdef} R^{cdef}-
         R_{cd}R^{cd}+\Box \hspace{-0.2ex} R \right) \nonumber \\
&& \;\;\;\;\;\;\;\;\;\;\;\;\;\;\;\;\;\;\;\;\;\;\;\;\;\;\;\;
\hspace{1ex}
      -\,2R^{acde}{R^b}_{cde}-2 R^{acbd}R_{cd}+4R^{ac}{R_c}^b
      -3 \hspace{0.2ex}\Box \hspace{-0.2ex} R^{ab}
  +\bigtriangledown^{a}\!\bigtriangledown^{b}\! \hspace{-0.2ex} R,
\label{D}
\end{eqnarray}
and
\be
\!\!\!\!\!\!\!\!\!\!\!\!\!\!\!\!\!\!\!\!\!\!\!\!\hspace{-6.3ex}
B^{ab} \equiv {1\over\sqrt{- g}}   \frac{\delta}{\delta g_{ab}}
 \int \! d^n x \,\sqrt{- g} \, R^2 
 = {1\over2}\, g^{ab} R^2-2 R R^{ab}+2 \bigtriangledown^{a}\!
   \bigtriangledown^{b}\hspace{-0.1ex} R  
   -2 g^{ab}\Box \hspace{-0.2ex} R.
\label{B in n}           
\ee
From equation 
(\ref{semiclassical eq in n}), after renormalizing the coupling
constants in order to 
eliminate the divergencies in 
$\mu^{-(n-4)}\langle \hat{T}_{n}^{ab}\rangle [g]$
in the limit $n\!\rightarrow \! 4$ and then taking this limit, we
will get the semiclassical Einstein equation in the physical spacetime
of four dimensions:
\be
{1\over 8 \pi G} \left( G^{ab}[g]+ \Lambda g^{ab} \right)-
2  \left( \alpha A^{ab}+\beta B^{ab} \right)\hspace{-0.3ex}[g]=
\left\langle\hat{T}_{R}^{ab}\right\rangle \![g].
\label{semiclassical Einstein eq}
\ee
In the last equation $1/G$, $\Lambda /G$, $\alpha$ and $\beta$ are
renormalized coupling constants, 
$\langle\hat{T}_{R}^{ab}\rangle [g]$ is the renormalized expectation
value of the stress-energy tensor operator, and we have used that, for
$n\!=\!4$, $D^{ab}=(3/2) A^{ab}$, being $A^{ab}$ the local curvature
tensor obtained by functional derivation with
respect to the metric of the action term corresponding 
to the Lagrangian density $C_{abcd}C^{abcd}$.


\subsection{\hspace{-2.5ex}. The semiclassical 
Einstein-Langevin equation}
\label{subsec:Einstein-Langevin}


According to the results of the previous section, assuming that some
suitably coarse-grained 
metric field satisfies the conditions for approximate decoherence
and that the approximations of subsection
\ref{sec:classicalization}\,\ref{subsec:effective eqs} 
are valid in a certain regime,
small deviations from a given solution $g_{ab}$ of the semiclassical
Einstein equation (\ref{semiclassical Einstein eq}) can be described
by linear stochastic perturbations $h_{ab}$ to that semiclassical
metric. These perturbations satisfy a Langevin equation of the
form (\ref{langevin eq}), which shall be called the semiclassical
Einstein-Langevin equation. Our next step will be to write the
semiclassical Einstein-Langevin equation in dimensional
regularization. Let us assume that $g_{ab}$ is a solution of
Eq.~(\ref{semiclassical eq in n}) in $n$ spacetime dimensions. The 
semiclassical Einstein-Langevin equation in dimensional
regularization has then the form
\begin{eqnarray}
{1\over 8 \pi G_{B}}\biggl(  G^{ab}_L[g\!+\!h]+ 
\Lambda_{B} \left(g^{ab}\!-\!h^{ab}\right) \biggr)\!
&\!\!\!\!\!\!\!-\!\!\!\!\!\!\!&
\!\left({4\over 3}\, \alpha_{B} D^{ab}_L
+ 2 \beta_{B} B^{ab}_L \right)\![g \!+\! h] = \mu^{-(n-4)} 
\left\langle \hat{T}_{n}^{ab}\right\rangle 
\!\!_{\mbox{}_{\scriptstyle L}}[g\!+\!h]\nonumber \\
&&\hspace{30 ex}\!+\,2 \mu^{-(n-4)} \xi_n^{ab}, 
\label{Einstein-Langevin eq in n} 
\end{eqnarray}
where $h_{ab}$ is a linear stochastic perturbation to 
$g_{ab}$, 
$h^{ab}\!\equiv\! g^{ac}g^{bd}h_{cd}$, that is, 
$g^{ab}\!-h^{ab}\!+0(h^2)$ is
the inverse of the metric $g_{ab}\!+\!h_{ab}$, and, as in the previous
section, we use a subindex ${\scriptstyle L}$ to denote an
expansion up to linear order in $h_{ab}$. 
In this equation, 
$\langle \hat{T}_{n}^{ab}\rangle [g+h]$ is 
the expectation value of 
$\hat{T}_{n}^{ab}[g\!+\!h]$
in the $n$-dimensional
spacetime generalization of the state described by
$\hat{\rho}[g+h]$,
and $\xi_n^{ab}$ is a Gaussian stochastic tensor
characterized by the correlators
\be
\left\langle\xi_n^{ab}(x) \right\rangle_{c}\!= 0, 
\hspace{10ex} 
\left\langle\xi_n^{ab}(x)\xi_n^{cd}(y) \right\rangle_{c}\!=
N_n^{abcd}[g](x,y),
\label{correlators in n}
\ee
with [see Eqs.~(\ref{gaussian correlators 2}) and
(\ref{C's 2})]
\be
2 N_n^{abcd}[g](x,y) \equiv 
\left. {1\over\sqrt{- g(x)}\sqrt{- g(y)} } \left[  
 \frac{\delta^2 \, {\rm Im}\, S^{\rm eff}_{\rm IF}[g^+,g^-]}
{\delta g^+_{ab}(x)\delta g^+_{cd}(y)}
- \frac{\delta^2 \, {\rm Im}\, S^{\rm eff}_{\rm IF}[g^+,g^-]}
{\delta g^+_{ab}(x)\delta g^-_{cd}(y)}
\right] \right|_{g^+=g^-=g}\!.
\label{noise in n}
\ee

We can write Eq.~(\ref{Einstein-Langevin eq in n}) in a more explicit
way by working out the expansion 
$\langle \hat{T}_{n}^{ab}\rangle 
\!_{\mbox{}_{\scriptstyle L}}[g+h]$. Since, from 
Eq.~(\ref{s-t expect value}), we have that
\be
\left\langle \hat{T}_n^{ab}(x) \right\rangle \![g+h]=
\left. {2\over\sqrt{-\det (g\!+\!h)(x)}} \, 
 \frac{\delta S^{\rm eff}_{\rm IF}
   [g\!+\!h^+,g\!+\!h^-]}{\delta h^+_{ab}(x)} 
 \right|_{h^+=h^-=h}\!,
\label{perturb s-t expect value}
\ee
this expansion can be obtained from an expansion of the influence
action $S^{\rm eff}_{\rm IF}[g+h^+,g+h^-]$ up to second order
in $h^{\pm}_{ab}$ (in this expansion, we can neglect boundary terms). 
At the same time, we can obtain a more explicit expression for the noise
kernel (\ref{noise in n}).
To perform this expansion for the
influence action, we have to compute the first and second order functional 
derivatives of $S^{\rm eff}_{\rm IF}[g^+,g^-]$
and then set $g^+_{ab}\!=\!g^-_{ab}\!=\!g_{ab}$.
If we do so using the path integral representation
(\ref{path integral}), we can interpret these derivatives as
expectation values of operators in the Heisenberg picture 
for a scalar field  
quantized in a classical spacetime background $({\cal M},g_{ab})$
as, for instance, in expression (\ref{s-t expect value}). 
The relevant second order derivatives are
\bea
\left. {1\over\sqrt{- g(x)}\sqrt{- g(y)} } \,   
 \frac{\delta^2 S^{\rm eff}_{\rm IF}[g^+,g^-]}
{\delta g^+_{ab}(x)\delta g^+_{cd}(y)}
 \right|_{g^+=g^-=g} \!\!
\!\!\!&=&\!\!\! 
-H_{\scriptscriptstyle \!{\rm S}_{\scriptstyle n}}^{abcd}[g](x,y)
-K_n^{abcd}[g](x,y)+
i N_n^{abcd}[g](x,y),      \nn \\
\left. {1\over\sqrt{- g(x)}\sqrt{- g(y)} } \, 
 \frac{\delta^2 S^{\rm eff}_{\rm IF}[g^+,g^-]}
{\delta g^+_{ab}(x)\delta g^-_{cd}(y)} 
 \right|_{g^+=g^-=g} \!\!
\!\!\!&=&\!\!\! 
-H_{\scriptscriptstyle \!{\rm A}_{\scriptstyle n}}^{abcd}
[g](x,y)
-i N_n^{abcd}[g](x,y),  
\label{derivatives}  
\eea
with
\bea
N_n^{abcd}[g](x,y) \!\!\!&= &\!\!\! 
{1\over 8}\, \biggl\langle  \biggl\{ 
 \hat{T}_n^{ab}(x)-
 \left\langle \hat{T}_n^{ab}(x) \right\rangle , \,
 \hat{T}_n^{cd}(y)- 
 \left\langle \hat{T}_n^{cd}(y)\right\rangle 
 \biggr\} \biggr\rangle [g],
\nn \\
H_{\scriptscriptstyle \!{\rm S}_{\scriptstyle n}}^{abcd}
[g](x,y)\!\!\! &= &\!\!\!
{1\over 4}\:{\rm Im} \left\langle {\rm T}^{\displaystyle \ast}\!\!
\left( \hat{T}_n^{ab}(x) \hat{T}_n^{cd}(y) 
\right) \right\rangle \![g],       \nn \\
H_{\scriptscriptstyle \!{\rm A}_{\scriptstyle n}}^{abcd}
[g](x,y) \!\!\!&= &\!\!\!
-{i\over 4}\, \left\langle {1\over 2} 
\left[ \hat{T}_n^{ab}(x), \, \hat{T}_n^{cd}(y)
\right] \right\rangle \![g],       \nn \\
K_n^{abcd}[g](x,y) \!\!\!&= &\!\!\!
\left. {-1\over\sqrt{- g(x)}\sqrt{- g(y)} } \, \left\langle 
\frac{\delta^2 S_m[g,\Phi_{n}]}{\delta g_{ab}(x)\delta g_{cd}(y)} 
\right|_{\Phi_{n}=\hat{\Phi}_{n}}\right\rangle \![g],
\label{kernels}  
\eea     
using again a symmetrical ordering (Weyl ordering) prescription
for the operators in the last of these expressions. All the
expectation values in these expressions are in the $n$-dimensional
spacetime generalization of the state described by
$\hat{\rho}[g]$.
In the above equations, $\{ \; , \: \}$ and  $[ \; , \: ]$ mean,
respectively, the anticommutator and the commutator, and we
use the symbol ${\rm T}^{\displaystyle \ast}$ to denote that, first,
we have to time order the field operators $\hat{\Phi}_{n}$ and then
apply the derivative operators that appear in each term 
of the product $T^{ab}(x) T^{cd}(y)$, where $T^{ab}$ is
the functional (\ref{class s-t}). For instance,
\be
{\rm T}^{\displaystyle \ast}\!\! \left(\hspace{-0.07ex} 
\bigtriangledown^{a}_{\!\!\! \mbox{}_{x}}
    \hspace{-0.1ex}\hat{\Phi}_{n}(x)\!
\bigtriangledown^{b}_{\!\!\! \mbox{}_{x}}\!\hat{\Phi}_{n}(x)\!
\bigtriangledown^{c}_{\!\!\! \mbox{}_{y}}\!\hat{\Phi}_{n}(y)\!
\bigtriangledown^{d}_{\!\!\! \mbox{}_{y}}\!\hat{\Phi}_{n}(y)\!
\right)\! =\!\!\!\!\lim_{ 
x_1,x_2 \rightarrow x_{\!\!\!\!\!\!\!\!\!\!\!\!\!\!\!\!\!\!\!\!\!
\!\!\!\!\!
\mbox{}_{\mbox{}_{\mbox{}_{\mbox{}_
{\mbox{}_{\scriptstyle x_3,x_4 \rightarrow y}}}}}} }\!\!\!
\bigtriangledown^{a}_{\!\!\! \mbox{}_{x_1}}\!\!\hspace{0.02ex}
\bigtriangledown^{b}_{\!\!\! \mbox{}_{x_2}}\!\!
\bigtriangledown^{c}_{\!\!\! \mbox{}_{x_3}}\!\!
\bigtriangledown^{d}_{\!\!\! \mbox{}_{x_4}}\!
{\rm T}\! \left(\hat{\Phi}_{n}(x_1)\hat{\Phi}_{n}(x_2)
\hat{\Phi}_{n}(x_3)\hat{\Phi}_{n}(x_4)  \right)\!,
\label{T star}
\ee
where ${\rm T}$ is the usual time ordering. 
Notice that all the kernels that appear 
in expressions (\ref{derivatives}) are real. 

In fact, from (\ref{kernels}), we see that the noise kernel
$N_n^{abcd}$, and also the kernel 
$H_{\scriptscriptstyle \!{\rm A}_{\scriptstyle n}}^{abcd}$, are
free of ultraviolet divergencies in the limit 
$n \!\rightarrow \!4$. This is because, for a linear quantum field,
the ultraviolet divergencies in $\left\langle\hat{T}_n^{ab}(x)
\hat{T}_n^{cd}(y)\right\rangle$ are the same ones as those
of $\left\langle\hat{T}_n^{ab}(x)\right\rangle
\left\langle\hat{T}_n^{cd}(y)\right\rangle$. 
Therefore, in the semiclassical Einstein-Langevin equation
(\ref{Einstein-Langevin eq in n}),
one can perform exactly the same renormalization procedure
as the one for the semiclassical Einstein equation 
(\ref{semiclassical eq in n}).
After this renormalization procedure, 
Eq.~(\ref{Einstein-Langevin eq in n}) will yield the semiclassical
Einstein-Langevin equation in the physical spacetime ($n\!=\!4$).
It can be written as 
\be
{1\over 8 \pi G} \Bigl( G^{ab}_L[g+h]+ 
\Lambda\left(g^{ab}-h^{ab}\right) \Bigr)- 
2 \left( \alpha A^{ab}_L+\beta B^{ab}_L \right)\hspace{-0.3ex}
[g+h]=\left\langle
\hat{T}_{R}^{ab}\right\rangle
\!\!_{\mbox{}_{\scriptstyle L}} [g+h] +2 \xi^{ab} , 
\label{Einstein-Langevin eq}
\ee  
being $\xi^{ab}$ is a Gaussian stochastic tensor with
\be
\left\langle\xi^{ab}(x) \right\rangle_c = 0,  
\hspace{6ex}
\left\langle\xi^{ab}(x)\xi^{cd}(y) \right\rangle_c = N^{abcd}[g](x,y),
\label{correlators}
\ee
where $N^{abcd} \equiv \lim_{n \rightarrow 4} \mu^{-2 (n-4)} 
N_n^{abcd}$. Notice from (\ref{kernels}) that
the noise kernel $N^{abcd}[g](x,y)$ gives a measure of the lowest
order fluctuations of the scalar field stress-energy tensor around its
expectation value. Thus, the
stochastic metric perturbations $h_{ab}$, solution of the
semiclassical Einstein-Langevin equation 
(\ref {Einstein-Langevin eq}), account for the back reaction of
such matter stress-energy fluctuations on the spacetime geometry. 
For a more detailed analysis of the semiclassical Einstein-Langevin
equation and some of its applications, see Ref.~\cite{mv98}.

Going back to the expressions in dimensional regularization, which may
be useful for calculational purposes, we can now write the 
expansion of the influence action around a given metric
$g_{ab}$. From (\ref{s-t expect value}) and
(\ref{derivatives}), taking into account that 
$S^{\rm eff}_{\rm IF}[g,g]=0$ and that 
$S^{\rm eff}_{\rm IF}[g^-,g^+]=
-S^{\rm eff {\displaystyle \ast}}_{\rm IF}[g^+,g^-]$, we get 
\bea
&&\hspace{-4ex}
S^{\rm eff}_{\rm IF}[g\!+\!h^+,g\!+\!h^-]
={1\over 2} \int\! d^nx\, \sqrt{- g(x)}\,
\left\langle \hat{T}_{n}^{ab}(x)\right\rangle \![g]
\left(h^+_{ab}(x)\!-\!h^-_{ab}(x) \right)  \nn \\
&&\hspace{-1.9ex}
-{1\over 2} \int\! d^nx\, d^ny \,\sqrt{- g(x)}\sqrt{- g(y)}
\left(H_{\scriptscriptstyle \!{\rm S}_{\scriptstyle n}}^{abcd}
[g](x,y)\!+\!K_n^{abcd}[g](x,y) \right)\!
\left(h^+_{ab}(x)h^+_{cd}(y)\!-\!h^-_{ab}(x)h^-_{cd}(y)  \right) 
\nn  \\
&&\hspace{-1.9ex}
-{1\over 2} \int\! d^nx\, d^ny\, \sqrt{- g(x)}\sqrt{- g(y)}\,
H_{\scriptscriptstyle \!{\rm A}_{\scriptstyle n}}^{abcd}
[g](x,y)
\left(h^+_{ab}(x)h^-_{cd}(y)\!-\!h^-_{ab}(x)h^+_{cd}(y)  \right) 
\nn  \\
&&\hspace{-1.9ex}
+{i\over 2} \int\! d^nx\, d^ny\, \sqrt{- g(x)}\sqrt{- g(y)}\,
N_n^{abcd}[g](x,y)
\left(h^+_{ab}(x)\!-\!h^-_{ab}(x) \right)
\left(h^+_{cd}(y)\!-\!h^-_{cd}(y) \right)+0(h^3).
\nn  \\
\mbox{}
\label{expansion 1}
\eea 
From (\ref{kernels}), it is easy to see that the
kernels satisfy the symmetry relations
\be
H_{\scriptscriptstyle \!{\rm S}_{\scriptstyle n}}^{abcd}(x,y)=
H_{\scriptscriptstyle \!{\rm S}_{\scriptstyle n}}^{cdab}(y,x), 
\hspace{3 ex} 
H_{\scriptscriptstyle \!{\rm A}_{\scriptstyle n}}^{abcd}(x,y)=
-H_{\scriptscriptstyle \!{\rm A}_{\scriptstyle n}}^{cdab}(y,x), 
\hspace{3 ex}
K_n^{abcd}(x,y) = K_n^{cdab}(y,x). 
\label{symmetries}
\ee
Using these relations, and defining
\be
H_n^{abcd}(x,y)\equiv 
H_{\scriptscriptstyle \!{\rm S}_{\scriptstyle n}}^{abcd}(x,y)
+H_{\scriptscriptstyle \!{\rm A}_{\scriptstyle n}}^{abcd}(x,y),
\label{H}
\ee
we can write the expansion (\ref{expansion 1}) as
\bea
S^{\rm eff}_{\rm IF}[g\!+\!h^+,\!\!\!\!&g&\!\!\!\!\!+h^-]
={1\over 2} \int\! d^nx\, \sqrt{- g(x)}\,
\left\langle \hat{T}_{n}^{ab}(x)\right\rangle \![g] \,
\left[h_{ab}(x) \right]  \nn \\
&&\hspace{-1ex}
-{1\over 2} \int\! d^nx\, d^ny\, \sqrt{- g(x)}\sqrt{- g(y)}\,
\left[h_{ab}(x)\right]
\left(H_n^{abcd}[g](x,y)\!
+\!K_n^{abcd}[g](x,y) \right)
\left\{ h_{cd}(y) \right\}  \nn  \\
&&\hspace{-1ex}
+{i\over 2} \int\! d^nx\, d^ny\, \sqrt{- g(x)}\sqrt{- g(y)}\,
\left[h_{ab}(x) \right]
N_n^{abcd}[g](x,y)
\left[h_{cd}(y) \right]+0(h^3),
\label{expansion 2}
\eea 
where we have used the notation
\be
\left[h_{ab}\right] \equiv h^+_{ab}\!-\!h^-_{ab},
\hspace{5 ex}
\left\{ h_{ab}\right\} \equiv h^+_{ab}\!+\!h^-_{ab}.
\label{notation}
\ee
Using this expansion and noting, from (\ref{kernels}), that  
\be
K_n^{abcd}[g](x,y)= -{1\over 4}  
\left\langle \hat{T}_{n}^{ab}(x)\right\rangle \![g]
{g^{cd}(x)\over\sqrt{- g(y)}}\, 
\delta^n(x\!-\!y)-{1\over 2}\,{1\over\sqrt{- g(y)}}
\left\langle \left.
\frac{\delta T^{ab}[g,\Phi_{n}](x)}{\delta g_{cd}(y)} 
\right|_{\Phi_{n}=\hat{\Phi}_{n}}\right\rangle \![g],
\label{K}
\ee
we get, from (\ref{perturb s-t expect value}),
\be
\left\langle \hat{T}_n^{ab}(x) \right\rangle 
 \!\!_{\mbox{}_{\scriptstyle L}} [g\!+\!h] =
\left\langle \hat{T}_n^{ab}(x) \hspace{-0.1ex}\right\rangle 
\![g] + \left\langle 
\hat{T}_n^{{\scriptscriptstyle (1)}\hspace{0.1ex} ab}
[g;h](x) \right\rangle \![g] -
2 \!\int\! d^ny \hspace{0.3ex} 
\sqrt{- g(y)} \hspace{0.3ex}  H_n^{abcd}[g](x,y) \hspace{0.2ex} 
h_{cd}(y),
\label{s-t expect value expansion}
\ee
where the operator 
$\hat{T}_n^{{\scriptscriptstyle (1)}\hspace{0.1ex} ab}$ is defined from
the term of first order in the expansion $T^{ab}_L[g+h,\Phi_{n}]$
as
\be
T^{ab}_L[g\!+\!h,\Phi_{n}]=T^{ab}[g,\Phi_{n}]+
T^{{\scriptscriptstyle (1)}\hspace{0.1ex} ab}[g,\Phi_{n};h],  
\hspace{5 ex}
\hat{T}_n^{{\scriptscriptstyle (1)}\hspace{0.1ex} ab}
[g;h]\equiv
T^{{\scriptscriptstyle (1)}\hspace{0.1ex} ab}[g,\hat{\Phi}_{n}[g];h], 
\label{T(1)}
\ee
using, as always, a Weyl ordering prescription for the operators in the
last definition. Note that the third term in the right hand side of
Eq.~(\ref{s-t expect value expansion}) is due to the dependence
on $h_{cd}$ of the field operator $\hat{\Phi}_{n}[g+h]$ and of the
dimensional regularized version of the density operator
$\hat{\rho}[g+h]$.

Substituting (\ref{s-t expect value expansion}) into 
(\ref{Einstein-Langevin eq in n}), and taking into account that
$g_{ab}$ satisfies the semiclassical Einstein equation
(\ref{semiclassical eq in n}), we can write the Einstein-Langevin
equation (\ref{Einstein-Langevin eq in n}) as
\bea
&&\hspace{-2ex}{1\over 8 \pi G_{B}}\left(
G^{{\scriptscriptstyle (1)}\hspace{0.1ex} ab}
[g;h](x)\!-\!
\Lambda_{B}\, h^{ab}(x) \right) -
{4\over 3}\, \alpha_{B} D^{{\scriptscriptstyle (1)}\hspace{0.1ex} ab}
[g;h](x)
-2\beta_{B} B^{{\scriptscriptstyle (1)}\hspace{0.1ex} ab}
[g;h](x)   \nn \\
&&\hspace{-2ex}- \mu^{-(n-4)} \hspace{-0.3ex}
\left\langle \hspace{-0.1ex}
\hat{T}_n^{{\scriptscriptstyle (1)}\hspace{0.1ex} ab}
[g;h](x) \hspace{-0.1ex}\right\rangle \![g]
\hspace{-0.2ex}+\hspace{-0.2ex}
2 \hspace{-0.2ex}\!\int\! d^ny\, \sqrt{- g(y)}\,\mu^{-(n-4)} 
H_n^{abcd}[g](x,y)\, h_{cd}(y)
\hspace{-0.2ex}=\hspace{-0.2ex} 
2 \mu^{-(n-4)} \xi_n^{ab}(x). \nn \\
\mbox{}
\label{Einstein-Langevin eq 2} 
\eea 
In the last
equation we have used the superindex ${\scriptstyle (1)}$ to denote
the terms of first order in the 
expansions $G^{ab}_L[g+h]$,
$D^{ab}_L[g+h]$ and $B^{ab}_L[g+h]$. Thus,
for instance, $G^{ab}_L[g+h]\!=\!G^{ab}[g]+
G^{{\scriptscriptstyle (1)}\hspace{0.1ex} ab}[g;h]$. 
The explicit expressions for the tensors 
$T^{{\scriptscriptstyle (1)}\hspace{0.1ex} ab}[g,\Phi_{n};h]$,
$G^{{\scriptscriptstyle (1)}\hspace{0.1ex} ab}[g;h]$,
$D^{{\scriptscriptstyle (1)}\hspace{0.1ex} ab}[g;h]$ and
$B^{{\scriptscriptstyle (1)}\hspace{0.1ex} ab}[g;h]$ are given in the
Appendix. 
From  
$T^{{\scriptscriptstyle (1)}\hspace{0.1ex} ab}[g,\Phi_{n};h]$, we can
write an explicit expression for the operator 
$\hat{T}_n^{{\scriptscriptstyle (1)}\hspace{0.1ex} ab}$.
Using the Klein-Gordon equation (\ref{Klein-Gordon in n}), and
expressions (\ref{regul s-t 2}) and (\ref{diff operator}) for the
stress-energy operator, we can write this operator as
\be
\hat{T}_n^{{\scriptscriptstyle (1)}\hspace{0.1ex} ab}
[g;h]=\left({1\over 2}\, g^{ab}h_{cd}-\delta^a_c h^b_d-
\delta^b_c h^a_d  \right) \hat{T}_{n}^{cd}[g]
+{\cal F}^{ab}[g;h]\, \hat{\Phi}_{n}^2[g],  
\label{T(1) operator}
\ee
where ${\cal F}^{ab}[g;h]$ is the differential operator
\bea
{\cal F}^{ab} \!\!\!\!&\equiv& \!\!\!\!\left(\xi\!-\!{1\over 4}\right)\! 
\left(h^{ab}\!-\!{1\over 2}\, g^{ab} h^c_c \right)\! \Box+
{\xi \over 2} \left[ 
\bigtriangledown^{c}\! \bigtriangledown^{a}\! h^b_c+
\bigtriangledown^{c}\! \bigtriangledown^{b}\! h^a_c- 
\Box h^{ab}-
\bigtriangledown^{a}\! \bigtriangledown^{b}\!  h^c_c-
g^{ab}\! \bigtriangledown^{c}\! \bigtriangledown^{d} h_{cd}
\right.   \nn \\
&&\!\!\!+\left. g^{ab} \Box h^c_c 
+\left( \bigtriangledown^{a} h^b_c+
\bigtriangledown^{b} h^a_c-\bigtriangledown_{\! c} 
\hspace{0.2ex} h^{ab}-
2 g^{ab}\! \bigtriangledown^{d}\! h_{cd} +
g^{ab}\! \bigtriangledown_{\! c} \! h^d_d
\right)\! \bigtriangledown^{c}
-g^{ab} h_{cd} \bigtriangledown^{c}\! \bigtriangledown^{d} 
\right],  \nn \\
\mbox{}
\label{diff operator F}
\eea
and it is understood that indices are raised
with the background inverse metric $g^{ab}$ and that all the
covariant derivatives are associated to the metric $g_{ab}$.
Substituting  expression (\ref{T(1) operator}) into 
Eq.~(\ref{Einstein-Langevin eq 2}), and using the semiclassical
equation 
(\ref{semiclassical eq in n}) to get an expression for 
$\mu^{-(n-4)}
\langle \hat{T}_{n}^{ab}\rangle [g]$, we can
finally write the semiclassical Einstein-Langevin equation in
dimensional regularization as
\bea
&&{1\over 8 \pi G_{B}}\Biggl[
G^{{\scriptscriptstyle (1)}\hspace{0.1ex} ab}\!-\!
{1\over 2}\, g^{ab} G^{cd} h_{cd}+ G^{ac} h^b_c+G^{bc} h^a_c+ 
\Lambda_{B} \left( h^{ab}\!-\!{1\over 2}\, g^{ab} h^c_c \right) 
\Biggr](x) -
{4\over 3}\, \alpha_{B}\biggl( D^{{\scriptscriptstyle
(1)}\hspace{0.1ex} ab}   \nn \\
&&\left. -{1\over 2}\, g^{ab} D^{cd} h_{cd}+ D^{ac} h^b_c+D^{bc} h^a_c
\right)\! (x)
-2\beta_{B}\left( B^{{\scriptscriptstyle (1)}\hspace{0.1ex} ab}\!-\!
{1\over 2}\, g^{ab} B^{cd} h_{cd}+ B^{ac} h^b_c+B^{bc} h^a_c 
\right)\! (x)   \nn \\
&&- \mu^{-(n-4)}\, {\cal F}^{ab}_x \!
\left\langle \hat{\Phi}_{n}^2(x) \right\rangle \![g]
+2 \!\int\! d^ny \, \sqrt{- g(y)}\, \mu^{-(n-4)} 
H_n^{abcd}[g](x,y)\, h_{cd}(y)
=2 \mu^{-(n-4)} \xi^{ab}_n(x),  \nn  \\
\mbox{}
\label{Einstein-Langevin eq 3} 
\eea 
where the tensors $G^{ab}$, $D^{ab}$ and
$B^{ab}$ are computed from the semiclassical metric $g_{ab}$,
and where we have omitted the functional dependence on $g_{ab}$ and
$h_{ab}$ in $G^{{\scriptscriptstyle (1)}\hspace{0.1ex} ab}$,
$D^{{\scriptscriptstyle (1)}\hspace{0.1ex} ab}$,
$B^{{\scriptscriptstyle (1)}\hspace{0.1ex} ab}$ and
${\cal F}^{ab}$ to simplify the notation. 
Notice that, in Eq.~(\ref{Einstein-Langevin eq 3}), 
all the ultraviolet divergencies in
the limit $n \!\rightarrow \!4$, which shall be removed by
renormalization of the coupling constants, are in 
$\left\langle \hat{\Phi}_{n}^2(x) \right\rangle$ and the
symmetric part 
$H_{\scriptscriptstyle \!{\rm S}_{\scriptstyle n}}^{abcd}(x,y)$ of the
kernel  $H_n^{abcd}(x,y)$, whereas, as we have pointed out above, the
kernels $N_n^{abcd}(x,y)$ and
$H_{\scriptscriptstyle \!{\rm A}_{\scriptstyle n}}^{abcd}(x,y)$ are
free of ultraviolet divergencies. 
Once we have performed such a renormalization procedure, setting
$n \!= \!4$ in this equation will yield the physical semiclassical
Einstein-Langevin equation, Eq.~(\ref{Einstein-Langevin eq}).
Note that, due to the presence of the kernel $H_n^{abcd}(x,y)$ in 
Eq.~(\ref{Einstein-Langevin eq 3}), such Einstein-Langevin equation
will be non-local in the metric perturbation.


\subsection{\hspace{-2.5ex}. Discussion}


We have seen that effective equations of motion for the metric field
of the form (\ref{semiclassical Einstein eq}) and 
(\ref{Einstein-Langevin eq}) follow from the local approximation 
(\ref{effective action ansatz}) for the
effective action describing the ``effective interaction'' of the
metric and the scalar field. 
A more realistic evaluation of this
effective action starting from a fundamental theory of quantum
gravity would certainly lead to some 
real and imaginary non-local terms in this action.
In some situations, the
contribution of these terms to the effective equations of motion for
the metric (note that they would also give some extra terms in the
semiclassical equation) might not be negligible and, in any case, one
would expect that their role in the decoherence mechanism for the metric
field would be important. This would
represent non trivial effects coming from the 
``high-momentum'' modes of quantum gravity, which are not part
of the gravitational field described by the classical stochastic metric
$g_{ab}+h_{ab}$, but which can be source of this gravitational
field in the same way as the matter fields. The contribution of these
neglected terms to the equations 
for the background metric $g_{ab}$ and for the stochastic
metric perturbation $h_{ab}$ would be similar to the contribution of the
scalar field through its stress-energy operator, but with this
operator replaced with some ``effective'' stress-energy operator of
such primordial ``high-momentum'' gravitational modes  
coupled to the scalar field. These equations would take the
form (\ref{semiclassical Einstein eq}) and 
(\ref{Einstein-Langevin eq}) 
only when the
effect of this ``effective'' stress-energy tensor on
the classical spacetime geometry can be neglected.
A way of partially modelizing this effect would consist on replacing
the stress-energy operator $\hat{T}_{n}^{ab}[g]$ by
$\hat{T}_{n}^{ab}[g]+\hat{t}_{n}^{ab}[g]$, where
$\hat{t}_{n}^{ab}[g]$ is the stress-energy tensor of gravitons
quantized in classical spacetime background $({\cal M},g_{ab})$
\cite{wald84}.

We end this paper with some comments on the relation between the 
semiclassical Einstein-Langevin equation (\ref{Einstein-Langevin eq})
and the Langevin-type equations for stochastic metric perturbations
recently derived in the literature 
\cite{calzettahu,humatacz,husinha,cv96,lomb-mazz,ccv97,campos-hu}.
In these previous derivations, one starts with the influence functional
(\ref{path integral}), with the state of the scalar field assumed to
be an ``in'' vacuum or 
an ``in'' thermal state, and computes explicitly the expansion for the
corresponding influence action around a specific metric background. 
One then applies the method of subsection 
\ref{sec:classicalization}\,\ref{subsec:quick method} to
derive a Langevin equation for the perturbations to this background. 
As we have seen in subsection 
\ref{sec:classicalization}\,\ref{subsec:quick method}, this method
yields the same equations as the one used in this section. 
However, in most of the previous derivations, one starts with a
``mini-superspace'' model and, thus, the metric perturbations are
assumed from the beginning to have a restrictive form. In those cases,
the derived Langevin equations do not correspond exactly to 
our equation, 
Eq.~(\ref{Einstein-Langevin eq}), but to a ``reduced''
version of this equation, in which only some components of the noise
kernel in Eq.~(\ref{correlators}) (or some particular combinations of
them) influence the dynamics of the metric perturbations. 
Only those
equations which have been derived starting from a completely general
form for the metric perturbations are actually particular cases,
computed explicitly, of the semiclassical Einstein-Langevin equation 
(\ref{Einstein-Langevin eq}) \cite{cv96,lomb-mazz,campos-hu}.


\section*{Acknowledgments}


We are grateful to Esteban Calzetta, Antonio Campos, Bei-Lok Hu and
Albert Roura for very helpful
suggestions and discussions.  This work has been
partially supported by the CICYT Research Project number
\mbox{AEN95-0590}, and the European Project number
\mbox{CI1-CT94-0004}.

\bigskip
\bigskip
\vspace{1ex}



{\noindent \Large \bf Appendix:
  Expansions around a background metric}
\appendix


\def\theequation{\Alph{section}.\arabic{equation}}
\def\thesubsection{\Alph{section}.\arabic{subsection}}
\setcounter{section}{1}
\setcounter{equation}{0}
\vspace{2ex}


For a metric of the form $\tilde{g}_{ab}\equiv g_{ab}+h_{ab}$, 
where $h_{ab}$ is a
small perturbation to a background metric $g_{ab}$, we list the
expansions of metric functionals around the background metric
up to linear order in the perturbation. In the following expressions,
all the tilded quantities refer to functionals constructed with the
metric $\tilde{g}_{ab}$, whereas that the analogous untilded ones are
constructed with the background metric $g_{ab}$. In particular,
$\taderiv_{\!a}$ and 
$\bigtriangledown_{\!a}$ are respectively
the covariant derivatives associated to the metric $\tilde{g}_{ab}$
and to the metric $g_{ab}$, and 
$\taderiv^a \equiv \tilde{g}^{\,ab} \taderiv_{\!b}$,
$\tBox \equiv \tderiv^a \! \taderiv_{\!a}$,
$\bigtriangledown^a \equiv g^{ab}\bigtriangledown_{\!b}$, 
$\Box \equiv \bigtriangledown^a \bigtriangledown_{\!a}$,
where $\tilde{g}^{\,ab}$ and $g^{ab}$ are respectively the inverses
of $\tilde{g}_{ab}$ and $g_{ab}$. We shall also raise indices in the
metric perturbation with the inverse background metric $g^{ab}$:
$h^a_b \equiv g^{ac}h_{cb}$ and $h^{ab} \equiv g^{ac} g^{bd}h_{cd}$.

\bea
&&\hspace{-3.6ex}\tilde{g}^{\,ab}=g^{ab}-h^{ab}+O(h^2), \\
&&\hspace{-3.6ex}
\sqrt{-\tilde{g}}= \sqrt{-g} \left( 1+{1 \over 2}\, h^a_a
+O(h^2) \right),  \\
&&\hspace{-3.6ex}\tilde{\Gamma}^{c}_{ab}=\Gamma^{c}_{ab}+{1 \over 2}\,
\bigl(\bigtriangledown_{\!a}h^c_b+\bigtriangledown_{\!b}h^c_a-
\bigtriangledown^{c}h_{ab} \bigr)+O(h^2). \\
&&\hspace{-3.6ex}\mbox{For a scalar function $f$},  \nn \\
&&\hspace{-3.6ex}\tderiv_{\!a}\! \tderiv_{\!b}\! f=
\bigtriangledown_{\!a}\! \hspace{-0.2ex}
\bigtriangledown_{\!b}\! f
-{1 \over 2} \bigtriangledown^{c}\!\! f \,
\bigl(\bigtriangledown_{\!a}h_{bc}+\bigtriangledown_{\!b}h_{ac}-
\bigtriangledown_{\!c}h_{ab}\bigr)+O(h^2), \\
&&\hspace{-3.6ex}\tBox f=\Box f
- \bigtriangledown^a \!\bigtriangledown^b\!\! f\: h_{ab}-
\bigtriangledown^a \!f \, \bigl(\bigtriangledown^b h_{ab}
-{1 \over 2}\bigtriangledown_{\!a}\!h^b_b
\hspace{0.2ex}\bigr)+O(h^2), \\
&&\hspace{-3.6ex}\tderiv^a \tderiv^b \!\!f=
\bigtriangledown^a \!\bigtriangledown^b \!\!f
-\bigtriangledown^a \!\bigtriangledown^c \!\!f \,h^b_c
-\bigtriangledown^b \!\bigtriangledown^c \!\!f \,h^a_c
-{1 \over 2} \bigtriangledown^{c}\!\! f\, 
\bigl(\bigtriangledown^a h^b_c
+\bigtriangledown^b h^a_c-\bigtriangledown_{\!c} h^{ab}
\hspace{0.2ex} \bigr)
 +O(h^2).  \\
&&\hspace{-3.6ex}\mbox{For a tensor $t^{ab}$},  \nn \\
&&\hspace{-3.6ex}\tBox t^{ab}=\Box t^{ab}\hspace{-0.2ex}
-\hspace{-0.2ex}\bigtriangledown^c\!\bigtriangledown^d \!t^{ab}
\hspace{0.2ex}h_{cd} 
+\bigl(g^{ae}\hspace{-0.2ex}\bigtriangledown^c\! t^{db}
\hspace{-0.2ex}+g^{be}\hspace{-0.2ex}\bigtriangledown^c\! t^{ad}
\hspace{-0.2ex}
-{1 \over 2}\,g^{cd}\hspace{-0.1ex}\bigtriangledown^e\! t^{ab}
\hspace{0.2ex}\bigr)\,
\bigl(\bigtriangledown_{\!c} h_{de}+\bigtriangledown_{\!d} h_{ce}
\hspace{-0.2ex}-\bigtriangledown_{\!e} h_{cd} \bigr) \nn \\
&&\hspace{-3.6ex}\hspace{7.3ex}
+\,{1 \over 2}\, 
\bigl( g^{ac}\hspace{0.2ex}t^{db}
+g^{bc\hspace{0.2ex}}t^{ad}\hspace{0.2ex}\bigr)\, 
\bigl( \bigtriangledown^e\!\bigtriangledown_{\!d} h_{ce}+
\Box h_{cd}- \bigtriangledown^e\!\bigtriangledown_{\!c} 
h_{de}\hspace{0.1ex}\bigr)
+O(h^2),  \\
&&\hspace{-3.6ex}\mbox{For the curvature tensors,}   \nn \\
&&\hspace{-3.6ex}\tilde{R}_{ab}=R_{ab}+{1 \over 2} \,
\bigl( \bigtriangledown^{c}\!\bigtriangledown_{\!a}\! h_{bc}
+\bigtriangledown^{c}\!\bigtriangledown_{\!b} h_{ac}
-\Box h_{ab}
-\bigtriangledown_{\!a}\!\bigtriangledown_{\!b} h^c_c 
\hspace{0.2ex} \bigr)
 +O(h^2), \\
&&\hspace{-3.6ex}
\tilde{R}^a_b=R^a_b-R^c_b\hspace{0.2ex}h^a_c+{1 \over 2} \,
\bigl( \bigtriangledown^{c}\!\bigtriangledown_{\!b}\! h^a_c
+\bigtriangledown^{c}\!\bigtriangledown^{a}\! h_{bc}
- \Box h^a_b
-\bigtriangledown_{\!b}\!\bigtriangledown^{a}\! h^c_c 
\hspace{0.2ex} \bigr)
 +O(h^2), \\
&&\hspace{-3.6ex}\tilde{R}=R-R^{ab}h_{ab}
+\bigtriangledown^a \!\bigtriangledown^b\! h_{ab}-\Box h^a_a
 +O(h^2), \\
&&\hspace{-3.6ex}
\tilde{R}^{ab}=R^{ab}\!-\!R^{ac} h^b_c\!-\!R^{bc} h^a_c 
+{1 \over 2}\, 
\bigl(\bigtriangledown^c \!\bigtriangledown^a\! h^b_c
+\!\bigtriangledown^c \!\bigtriangledown^b\! h^a_c
-\Box h^{ab}
-\!\bigtriangledown^a \!\bigtriangledown^b\! h^c_c
\hspace{0.2ex} \bigr)
+O(h^2), \\
&&\hspace{-3.6ex}\tilde{G}^{ab}=G^{ab}
+G^{{\scriptscriptstyle (1)}\hspace{0.1ex} ab}+O(h^2), 
\hspace{3ex} \mbox{with}  \nn \\
&&\hspace{-3.6ex}
G^{{\scriptscriptstyle (1)}\hspace{0.1ex} ab}=
-R^{ac} h^b_c\!-\!R^{bc} h^a_c
+{1 \over 2} \, \bigl[\hspace{0.1ex} R\, h^{ab}
\!+\! g^{ab} R^{cd} h_{cd} 
+\!\bigtriangledown^c \!\bigtriangledown^a\! h^b_c
+\!\bigtriangledown^c \!\bigtriangledown^b\! h^a_c
-\Box h^{ab}
\!-\!\bigtriangledown^a \!\bigtriangledown^b\! h^c_c
\nn \\
&&\hspace{-3.6ex}\hspace{8.55ex}  
+\,g^{ab}\, (\Box h^c_c
-\!\bigtriangledown^c \!\bigtriangledown^d\! h_{cd}) 
\hspace{0.1ex}\bigr]
=-G^{ac} h^b_c-G^{bc} h^a_c  
+{1 \over 2} \, 
\bigl[\hspace{0.1ex}- R\, h^{ab}+g^{ab} R^{cd} h_{cd}
\nn \\
&&\hspace{-3.6ex}\hspace{8.55ex}
+\!\bigtriangledown^c \!\bigtriangledown^a h^b_c
+\!\bigtriangledown^c \!\bigtriangledown^b\! h^a_c
-\Box h^{ab}
-\!\bigtriangledown^a \!\bigtriangledown^b\! h^c_c
+g^{ab}\, (\Box h^c_c
-\!\bigtriangledown^c \!\bigtriangledown^d\! h_{cd}) 
\hspace{0.1ex}\bigr],   \\
&&\hspace{-3.6ex}
\tilde{R}^a_{\;\:bcd}=R^a_{\;\:bcd}+{1 \over 2}\, \bigl( 
\bigtriangledown_{\!c}\!\bigtriangledown_{\!b}\! h^a_d
+\bigtriangledown_{\!c}\!\bigtriangledown_{\!d}\! h^a_b
+\bigtriangledown_{\!d}\!\bigtriangledown^a\! h_{bc}
-\bigtriangledown_{\!c}\!\bigtriangledown^a\! h_{bd}
-\bigtriangledown_{\!d}\!\bigtriangledown_{\!b}\! h^a_c
\nn \\
&&\hspace{-3.6ex}\hspace{18.9ex}
-\bigtriangledown_{\!d}\!\bigtriangledown_{\!c} h^a_b
\hspace{0.2ex} \bigr)+O(h^2),  \\
&&\hspace{-3.6ex}\tilde{R}_{abcd}=R_{abcd}+
{1 \over 2}\, \bigl(
R^e_{\;\:bcd}\hspace{0.2ex}h_{ae}
+R^{\;\;e}_{a\;\;cd}\hspace{0.2ex}h_{be} \hspace{0.2ex} \bigr)
+{1 \over 2}\, \bigl( 
\bigtriangledown_{\!c}\!\bigtriangledown_{\!b}\! h_{ad}
+\bigtriangledown_{\!d}\!\bigtriangledown_{\!a}\! h_{bc}
-\bigtriangledown_{\!c}\!\bigtriangledown_{\!a}\! h_{bd} \nn \\
&&\hspace{-3.6ex}\hspace{45.2ex}
-\bigtriangledown_{\!d}\!\bigtriangledown_{\!b}
\hspace{0.1ex} h_{ac}
\hspace{0.2ex} \bigr)+O(h^2),  \\
&&\hspace{-3.6ex}\tilde{R}^{abcd}=R^{abcd}
-{1 \over 2}\, \bigl(2 R^{abce}\hspace{0.2ex}h^d_e
+2 R^{abed}\hspace{0.2ex}h^c_e
+R^{aecd}\hspace{0.2ex}h^b_e
+R^{ebcd}\hspace{0.2ex}h^a_e
\hspace{0.2ex} \bigr) \nn \\
&&\hspace{-3.6ex}\hspace{8.3ex}
+\,{1 \over 2}\, \bigl(
\bigtriangledown^c \!\bigtriangledown^b \! h^{ad}
+\bigtriangledown^d \!\bigtriangledown^a \! h^{bc}
-\bigtriangledown^c \!\bigtriangledown^a \! h^{bd}
-\bigtriangledown^d \!\bigtriangledown^b \! h^{ac} 
\hspace{0.2ex} \bigr)+O(h^2), \\
&&\hspace{-3.6ex}\tderiv^a \tderiv^b \! \tilde{R}=
\bigtriangledown^a \!\bigtriangledown^b\! R
-\bigtriangledown^a \!\bigtriangledown^b\! (R^{cd}h_{cd})
+\bigtriangledown^a \!\bigtriangledown^b\!
\bigtriangledown^c \!\bigtriangledown^d\! h_{cd}
-\bigtriangledown^a \!\bigtriangledown^b \!\Box h^c_c
-\bigtriangledown^a \!\bigtriangledown^c \!
\hspace{-0.1ex} R\: h^b_c \nn \\ 
&&\hspace{-3.6ex}\hspace{11ex}
-\bigtriangledown^b \!\bigtriangledown^c  R\: h^a_c
-{1 \over 2}  \bigtriangledown^c \!R \: 
\bigl(\bigtriangledown^a  h^b_c+\bigtriangledown^b  h^a_c
-\bigtriangledown_{\!c}h^{ab}\hspace{0.2ex} \bigr)+O(h^2),  \\
&&\hspace{-3.6ex}
\tBox \tilde{R}=\Box R-\Box\hspace{0.2ex} (R^{ab}h_{ab})
+\Box \!\bigtriangledown^a \!\bigtriangledown^b h_{ab}
-\Box^2 h^a_a
- \bigtriangledown^a \!\bigtriangledown^b\!\! R\: h_{ab} \nn \\
&&\hspace{-3.6ex}\hspace{6.5ex}
-\bigtriangledown^a \hspace{-0.2ex}\!R \: 
\bigl(\bigtriangledown^b h_{ab}
-{1 \over 2}\bigtriangledown_{\!a}\!h^b_b\hspace{0.2ex} \bigr)
+O(h^2), \\ 
&&\hspace{-3.6ex}\tBox \tilde{R}^{ab}=\Box R^{ab}
-\Box (R^{ac}\hspace{0.1ex}h^b_c+R^{bc}\hspace{0.1ex}h^a_c)
-\bigtriangledown^c \!\bigtriangledown^d\! R^{ab}\hspace{0.2ex} h_{cd}
-\bigtriangledown^c  R^{ab} \, \bigl(
\bigtriangledown^d h_{cd}
-{1 \over 2} \bigtriangledown_{\!c} \! h^d_d
\hspace{0.2ex} \bigr) \nn \\
&&\hspace{-3.6ex}\hspace{8.2ex}
+\bigtriangledown^c\!  R^{ad} \, \bigl(
\bigtriangledown_{\!c}\hspace{0.2ex} h^b_d
+\bigtriangledown_{\!d}\hspace{0.2ex} h^b_c
-\bigtriangledown^b h_{cd} \hspace{0.2ex} \bigr)
+\bigtriangledown^c  R^{bd} \, \bigl(
\bigtriangledown_{\!c}\hspace{0.2ex} h^a_d
+\bigtriangledown_{\!d}\hspace{0.2ex} h^a_c
-\bigtriangledown^a h_{cd} \hspace{0.2ex} \bigr) \nn \\
&&\hspace{-3.6ex}\hspace{8.2ex}
+\,{1 \over 2} \, R^{ac}\, \bigl(
\bigtriangledown^d \!\bigtriangledown_{\!c}\! h^b_d+
\Box h^b_c-\bigtriangledown^d \!\bigtriangledown^b\! h_{cd}
 \hspace{0.2ex} \bigr)
+{1 \over 2} \, R^{bc}\, \bigl(
\bigtriangledown^d \!\bigtriangledown_{\!c}\! h^a_d+
\Box h^a_c-\bigtriangledown^d \!\bigtriangledown^a\! h_{cd}
 \hspace{0.2ex} \bigr) \nn \\
&&\hspace{-3.6ex}\hspace{8.2ex}
+\,{1 \over 2} \, \bigl( 
\Box \!\bigtriangledown^c \!\bigtriangledown^a h^b_c
+\Box \!\bigtriangledown^c \!\bigtriangledown^b h^a_c
-\Box^2 h^{ab}\hspace{-0.2ex}
-\Box \!\bigtriangledown^a \!\bigtriangledown^b h^c_c
 \hspace{0.2ex} \bigr)+O(h^2),  \\
&&\hspace{-3.6ex}\tilde{R}^2=R^2-2R\hspace{0.1ex} R^{ab}h_{ab}
+2R\hspace{-0.1ex}\bigtriangledown^a \!\bigtriangledown^b h_{ab}
-2R \: \Box h^a_a +O(h^2), \\
&&\hspace{-3.6ex}
\tilde{R}\hspace{0.1ex}\tilde{R}^{ab}=R\hspace{0.1ex} R^{ab}
-R\hspace{0.1ex} R^{ac}\hspace{0.1ex} h^b_c
-R\hspace{0.1ex} R^{bc}\hspace{0.1ex} h^a_c
-R^{ab}R^{cd}h_{cd}
+{1 \over 2}\,R\: \bigl( 
\bigtriangledown^c \! \bigtriangledown^a\! h^b_c
+\bigtriangledown^c \! \bigtriangledown^b\! h^a_c  \nn \\
&&\hspace{-3.6ex}\hspace{11ex}
-\,\Box h^{ab}\!
-\bigtriangledown^a \!\bigtriangledown^b\! h^c_c 
\hspace{0.2ex} \bigr)
+R^{ab} \, \bigl(
\bigtriangledown^c \!\bigtriangledown^d\! h_{cd}
-\Box h^c_c \hspace{0.2ex} \bigr) +O(h^2),  \\
&&\hspace{-3.6ex}\tilde{R}^{ab}\tilde{R}_{ab}=R^{ab} R_{ab}
-2R^{ab}R_a^c\hspace{0.2ex}h_{bc}+
R^{ab} \hspace{0.2ex}
\bigl(2 \bigtriangledown^{c}\! \bigtriangledown_{\!a} h_{bc}
-\Box h_{ab}
-\bigtriangledown_{\!a}\!\bigtriangledown_{\!b}\! h^c_c
\hspace{0.2ex} \bigr)+ O(h^2), \\
&&\hspace{-3.6ex}
\tilde{R}^{ac}\tilde{R}^b_c=R^{ac}R^b_c-R^{ac}R^{bd}h_{cd}
-R^{cd} \, \bigl( R^a_c\hspace{0.1ex} h^b_d
+R^b_c\hspace{0.1ex} h^a_d \hspace{0.2ex} \bigr)
+{1 \over 2}\,R^{ac}\, \bigl( 
\bigtriangledown^d\! \bigtriangledown_{\!c}\! h^b_d
+\bigtriangledown^d\! \bigtriangledown^b\! h_{cd}
\nn \\
&&\hspace{-3.6ex}\hspace{9ex}
-\hspace{0.1ex}\Box h^b_c 
-\!\bigtriangledown_{\!c}\!\bigtriangledown^b\! h^d_d 
\hspace{0.2ex} \bigr) 
+{1 \over 2}\,R^{bc}\, \bigl( 
\bigtriangledown^d\! \bigtriangledown_{\!c}\! h^a_d
+\hspace{-0.1ex}\bigtriangledown^d\! \bigtriangledown^a\! h_{cd}
-\hspace{-0.1ex}\Box h^a_c 
-\!\bigtriangledown_{\!c}\!\bigtriangledown^a\! h^d_d 
\hspace{0.2ex} \bigr)
\hspace{-0.2ex}+\hspace{-0.2ex} O(h^2),  \\
&&\hspace{-3.6ex}
\tilde{R}^{abcd} \tilde{R}_{abcd}=R^{abcd}R_{abcd}
-2 R^{abcd}R_{abce} \hspace{0.3ex}h^e_d
+4 R^{abcd}\hspace{-0.1ex}
\bigtriangledown_{\!c}\!\bigtriangledown_{\!b}
\hspace{0.1ex} h_{ad}
+O(h^2), \\
&&\hspace{-3.6ex}\tilde{R}^{acbd}\tilde{R}_{cd}=R^{acbd} R_{cd}
+{1 \over 2}\,R_{cd} \, \bigl(
R^{acde}\hspace{0.3ex} h^b_e
+R^{bcde}\hspace{0.3ex} h^a_e
-2 R^{acbe}\hspace{0.3ex} h^d_e
-2 R^{bcae}\hspace{0.3ex} h^d_e
\hspace{0.2ex} \bigr)  \nn \\
&&\hspace{-3.6ex}\hspace{11.8ex}
+\,{1 \over 2}\, R^{acbd} \, \bigl( 
\bigtriangledown^e \! \bigtriangledown_{\!c}\! h_{de}
+\bigtriangledown^e \! \bigtriangledown_{\!d}\! h_{ce}
-\Box h_{cd}
-\bigtriangledown_{\!c}\!\bigtriangledown_{\!d}\! h^e_e
\hspace{0.2ex} \bigr)
-{1 \over 4}\, R_{cd} \, \bigl(
2 \bigtriangledown^c \! \bigtriangledown^d  h^{ab} \nn \\
&&\hspace{-3.6ex}\hspace{22.5ex}
+\bigtriangledown^a \! \bigtriangledown^b  h^{cd}
+\bigtriangledown^b \! \bigtriangledown^a \! h^{cd}
-2 \bigtriangledown^a \! \bigtriangledown^c  h^{bd}
-2 \bigtriangledown^b \! \bigtriangledown^c  h^{ad}
\hspace{0.2ex} \bigr)+O(h^2),  \\
&&\hspace{-3.6ex}
\tilde{R}^{acde} \tilde{R}^b_{\;\:cde}=R^{acde} R^b_{\;\:cde}
-{1 \over 2}\, \bigl( R^{acde}R^f_{\;\:cde}\hspace{0.3ex}h^b_f
+R^{bcde}R^f_{\;\:cde}\hspace{0.3ex}h^a_f\hspace{0.2ex} \bigr)
-2 R^{acde} R^b_{\;\:cdf}\hspace{0.3ex} h^f_e \nn \\
&&\hspace{-3.6ex}\hspace{13.2ex}
+\,{1 \over 2}\, R^{acde} \, \bigl( 
\bigtriangledown_{\!d}\!\bigtriangledown_{\!c}\! h^b_e
+\bigtriangledown_{\!e}\!\bigtriangledown^b \! h_{cd}
-\bigtriangledown_{\!e}\!\bigtriangledown_{\!c}\! h^b_d
-\bigtriangledown_{\!d}\!\bigtriangledown^b \! h_{ce}
\hspace{0.2ex} \bigr) \nn \\
&&\hspace{-3.6ex}\hspace{13.2ex}
+\,{1 \over 2}\, R^{bcde} \, \bigl( 
\bigtriangledown_{\!d}\!\bigtriangledown_{\!c}\! h^a_e
+\bigtriangledown_{\!e}\!\bigtriangledown^a \! h_{cd}
-\bigtriangledown_{\!e}\!\bigtriangledown_{\!c}\! h^a_d
-\bigtriangledown_{\!d}\!\bigtriangledown^a \! h_{ce}
\hspace{0.2ex} \bigr)+O(h^2),  \\
&&\hspace{-3.6ex}
\tilde{B}^{ab}=B^{ab}+B^{{\scriptscriptstyle (1)}\hspace{0.1ex} ab}
+O(h^2), \hspace{3ex} \mbox{with}  \nn \\
&&\hspace{-3.6ex}B^{{\scriptscriptstyle (1)}\hspace{0.1ex} ab}=
-{1 \over 2}\,R^2 h^{ab}
-g^{ab}R\hspace{0.1ex}R^{cd}h_{cd}
+2 R \, \bigl(R^{ac}h^b_c+R^{bc}h^a_c \hspace{0.2ex} \bigr)
+ R\hspace{-0.2ex}\bigtriangledown^a \!\bigtriangledown^b h^c_c
+2 R^{ab}\, \bigl(R^{cd}h_{cd}+\Box h^c_c
  \nn \\
&&\hspace{-3.6ex}\hspace{2ex}
-\bigtriangledown^c \! \bigtriangledown^d h_{cd} \bigr) 
+g^{ab}\hspace{-0.1ex} 
 \bigtriangledown^c \! \bigtriangledown^d R\, h_{cd}
+2\, \Box R \: h^{ab} 
-2 \hspace{-0.2ex}\bigtriangledown^c \!\bigtriangledown^a R\: h^b_c 
-2 \hspace{-0.2ex}\bigtriangledown^c \!\bigtriangledown^b R\: h^a_c
+g^{ab}\hspace{-0.2ex}
\bigtriangledown^c \! \bigtriangledown^d 
(R\hspace{0.1ex} h_{cd}) 
 \nn \\
&&\hspace{-3.6ex}\hspace{2ex}
-\bigtriangledown^c 
\bigl[\hspace{0.1ex} R\: \bigl( 
g^{ab}\hspace{-0.2ex} \bigtriangledown_{\!c}\!h^d_d
+\bigtriangledown^a h^b_c+\bigtriangledown^b h^a_c
-\bigtriangledown_{\!c} h^{ab}\hspace{0.2ex} \bigr)
\hspace{0.1ex}\bigr]
+2\, g^{ab} \,\Box\hspace{0.2ex} \bigl(R^{cd}h_{cd}+\Box h^c_c
-\bigtriangledown^c \! \bigtriangledown^d \!h_{cd} \bigr) 
 \nn \\
&&\hspace{-3.6ex}\hspace{2ex}
-\,2 \hspace{-0.2ex}\bigtriangledown^a \!\bigtriangledown^b
\bigl(R^{cd}h_{cd}+\Box h^c_c
-\bigtriangledown^c \! \bigtriangledown^d \!h_{cd} \bigr), \\
&&\hspace{-3.6ex}
\tilde{D}^{ab}=D^{ab}+D^{{\scriptscriptstyle (1)}\hspace{0.1ex} ab} 
+O(h^2), \hspace{3ex} \mbox{with}  \nn \\
&&\hspace{-3.6ex}D^{{\scriptscriptstyle (1)}\hspace{0.1ex} ab}=
{1 \over 2}\, \bigl(R^{cd}R_{cd}-R^{cdef}R_{cdef}\bigr) 
\hspace{0.2ex} h^{ab}
+2 R_{cdef}\, \bigl(R^{acde} h^{bf}+R^{bcde} h^{af}
\hspace{0.2ex}\bigr)
-R_{cd} \,\bigl(4 R^{ac}h^{bd} \nn \\
&&\hspace{-3.6ex}\hspace{2ex}
+\,4 R^{bc}h^{ad}+R^{acde} h^b_e+R^{bcde} h^a_e
-2 R^{acbe} h^d_e-2 R^{bcae} h^d_e \hspace{0.2ex}\bigr)
+g^{ab}\,\bigl(R^{cf}R_{cg}-R^{cdef}R_{cdeg} \bigr)
\hspace{0.2ex} h^g_f \nn \\
&&\hspace{-3.6ex}\hspace{2ex}
-\,4 R^{ac}R^{bd} h_{cd}
+4 R^{acde} R^b_{\;\:cdf} \hspace{0.2ex} h^f_e
+{1 \over 2}\, R^{cd}\, \bigl(\hspace{0.1ex} 
2\hspace{-0.1ex} \bigtriangledown_{\!c}\!\bigtriangledown_{\!d}
  \hspace{0.1ex} h^{ab} 
+\bigtriangledown^a \!\bigtriangledown^b \! h_{cd}
+\bigtriangledown^b \!\bigtriangledown^a \! h_{cd}  \nn \\
&&\hspace{-3.6ex}\hspace{2ex}
-\,2 \bigtriangledown^a \!\bigtriangledown_{\!c}\hspace{0.1ex} h^b_d 
-2 \bigtriangledown^b \!\bigtriangledown_{\!c}\hspace{0.1ex} h^a_d
\hspace{0.2ex} \bigr)
+{1 \over 2}\, R^{ac}\, \bigl(
\bigtriangledown^d \!\bigtriangledown_{\!c}\! h^b_d
-7\hspace{0.2ex} \Box h^b_c
+7 \bigtriangledown^d \!\bigtriangledown^b  h_{cd}
-4 \bigtriangledown_{\!c}\! \bigtriangledown^b  h^d_d 
\hspace{0.2ex} \bigr)  \nn \\
&&\hspace{-3.6ex}\hspace{2ex}
+\,{1 \over 2}\, R^{bc}\, \bigl(
\bigtriangledown^d \!\bigtriangledown_{\!c}\hspace{-0.2ex} h^a_d
-7\hspace{0.2ex} \Box h^a_c
+7 \bigtriangledown^d \!\bigtriangledown^a  h_{cd}
-4 \bigtriangledown_{\!c}\! \bigtriangledown^a  h^d_d 
\hspace{0.2ex} \bigr)
-R^{acde} \, \bigl(
\bigtriangledown_{\!d}\!\bigtriangledown_{\!c}\! h^b_e
+\bigtriangledown_{\!e}\!\bigtriangledown^b \! h_{cd} \nn \\
&&\hspace{-3.6ex}\hspace{2ex}
-\bigtriangledown_{\!d}\!\bigtriangledown^b h_{ce}
-\bigtriangledown_{\!e}\!\bigtriangledown_{\!c}\! h^b_d
\hspace{0.2ex} \bigr)
-R^{bcde} \, \bigl(
\bigtriangledown_{\!d}\!\bigtriangledown_{\!c}\! h^a_e
+\bigtriangledown_{\!e}\!\bigtriangledown^a \! h_{cd}
-\bigtriangledown_{\!d}\!\bigtriangledown^a\! h_{ce}
-\bigtriangledown_{\!e}\!\bigtriangledown_{\!c}\! h^a_d
\hspace{0.2ex} \bigr) \nn \\
&&\hspace{-3.6ex}\hspace{2ex}
-\,{1 \over 2}\, g^{ab} R^{cd}\, \bigl(\hspace{0.1ex} 
2\hspace{-0.1ex} 
\bigtriangledown^e \!\bigtriangledown_{\!c} h_{de} 
-\Box h_{cd}
-\bigtriangledown_{\!c}\!\bigtriangledown_{\!d}\! h^e_e
\hspace{0.2ex} \bigr)
+2\hspace{0.2ex} g^{ab} R^{cdef} \hspace{-0.2ex}
\bigtriangledown_{\!e}\!\bigtriangledown_{\!d}\hspace{0.1ex} h_{cf}
-R^{acbd} \, \bigl( 
\bigtriangledown^e \!\bigtriangledown_{\!c}\! h_{de} \nn \\
&&\hspace{-3.6ex}\hspace{2ex}
+\bigtriangledown^e \hspace{-0.1ex}\!\bigtriangledown_{\!d}
\hspace{0.1ex} h_{ce} 
-\Box h_{cd}
-\bigtriangledown_{\!c}\!\bigtriangledown_{\!d}\! h^e_e
\hspace{0.2ex} \bigr)
+{1 \over 2} \bigtriangledown^c\!\hspace{-0.2ex} R 
\: \bigl( 
\bigtriangledown_{\!c}\hspace{0.2ex} h^{ab}
-\bigtriangledown^a h^b_c
-\bigtriangledown^b h^a_c \hspace{0.2ex} \bigr)
-3 \bigtriangledown^c\!\hspace{-0.2ex} R ^{ad}
\, \bigl( 
\bigtriangledown_{\!c}\hspace{0.2ex} h^b_d  \nn \\
&&\hspace{-3.6ex}\hspace{2ex}
+\bigtriangledown_{\!d}\! h^b_c
-\hspace{-0.2ex}\bigtriangledown^b h_{cd} \hspace{0.2ex} \bigr)
-3 \bigtriangledown^c\!\hspace{-0.3ex} R ^{bd}
\, \bigl( 
\bigtriangledown_{\!c}\hspace{0.2ex} h^a_d
+\hspace{-0.1ex}\bigtriangledown_{\!d}\hspace{0.2ex} h^a_c
-\hspace{-0.2ex}\bigtriangledown^a h_{cd} \hspace{0.2ex} \bigr)
-{1 \over 4}\, \bigl( 
g^{ab}\hspace{-0.1ex} \bigtriangledown^c\!\hspace{-0.3ex} R
-6\hspace{-0.1ex} \bigtriangledown^c\!\hspace{-0.3ex} R^{ab} \bigr)
\hspace{0.2ex} \bigl( 2 \bigtriangledown^d \! h_{cd} \nn \\
&&\hspace{-3.6ex}\hspace{2ex}
-\bigtriangledown_{\!c}\hspace{-0.2ex} h^d_d \hspace{0.2ex} \bigr)
-{1 \over 2} \; \Box R\: h^{ab}
-\bigtriangledown^a \! \bigtriangledown^c\!\hspace{-0.2ex} R 
 \: h^b_c
-\bigtriangledown^b \! \bigtriangledown^c\!\hspace{-0.2ex} R 
 \: h^a_c
-\,{1 \over 2}\, g^{ab} 
\bigtriangledown^c \! \bigtriangledown^d \hspace{-0.1ex} R
 \: h_{cd}
+3 \bigtriangledown^c \! \bigtriangledown^d \hspace{-0.1ex} R^{ab}
 \: h_{cd} \nn  \\
&&\hspace{-3.6ex}\hspace{2ex}
+\, {3 \over 2} \: \Box \hspace{0.2ex} \bigl( 
2 R^{ac} h^b_c+ 2 R^{bc} h^a_c 
+\Box h^{ab}
\hspace{-0.1ex}+\hspace{-0.1ex}
\bigtriangledown^a \!\bigtriangledown^b \! h^c_c
\hspace{-0.1ex}-\hspace{-0.1ex}
\bigtriangledown^c\! \bigtriangledown^a \!h^b_c
\hspace{-0.1ex}-\hspace{-0.1ex}
\bigtriangledown^c\! \bigtriangledown^b \! h^a_c
\hspace{0.2ex}\bigr)
-{1 \over 2}\: g^{ab} \, \Box\hspace{0.2ex} \bigl(
R^{cd}h_{cd}
+\Box h^c_c \nn  \\
&&\hspace{-3.6ex}\hspace{2ex}
-\bigtriangledown^c \!\bigtriangledown^d h_{cd} 
\hspace{0.2ex}\bigr)
-\bigtriangledown^a \!\bigtriangledown^b \hspace{-0.3ex}
\bigl(
R^{cd}h_{cd}
+\Box h^c_c
-\bigtriangledown^c \!\bigtriangledown^d \! h_{cd} 
\hspace{0.2ex}\bigr). \nn \\
&&\hspace{-3.6ex}\mbox{}  \\
&&\hspace{-3.6ex}
\mbox{For the stress-energy tensor functional},  \nn \\ 
&&\hspace{-3.6ex}
T^{ab}[g,\Phi_{n}]\equiv \bigtriangledown^a \Phi_{n} \!
\bigtriangledown^b \!\Phi_{n}- {1\over 2}\, g^{ab}\hspace{-0.1ex} 
\bigtriangledown^{c}\!\Phi_{n}\! \bigtriangledown_{\!c}\!\Phi_{n} 
-{1\over 2}\, g^{ab}\hspace{0.2ex} m^2 \Phi_{n}^2 
+\xi \left( g^{ab} \Box
-\bigtriangledown^a \!\bigtriangledown^b
+G^{ab} \right)  \Phi_{n}^2,  \nn \\
&&\hspace{-3.6ex}T^{ab}[\tilde{g},\Phi_{n}]=T^{ab}[g,\Phi_{n}]+
T^{{\scriptscriptstyle (1)}\hspace{0.1ex} ab}[g,\Phi_{n};h]
+0(h^2), \hspace{3ex} \mbox{with}  \nn \\
&&\hspace{-3.6ex}
T^{{\scriptscriptstyle (1)}\hspace{0.1ex} ab}[g,\Phi_{n};h]=
-T^{ac}[g,\Phi_{n}]\, h^b_c-T^{bc}[g,\Phi_{n}]\, h^a_c
-{1 \over 2}\, \bigl( 
\bigtriangledown^{c}\Phi_{n}\! \bigtriangledown_{\!c}\!\Phi_{n}
+m^2 \Phi_{n}^2 \hspace{0.1ex}\bigr) \, h^{ab} \nn \\
&&\hspace{-3.6ex}\hspace{2ex}
+\,{1 \over 2}\, g^{ab} \bigtriangledown^c\! \Phi_{n}\!
\bigtriangledown^d\! \Phi_{n} \, h_{cd} 
+{\xi \over 2} \left[ -R\, h^{ab}+g^{ab} R^{cd}h_{cd}
+\bigtriangledown^c \!\bigtriangledown^a \! h^b_c
+\bigtriangledown^c \!\bigtriangledown^b \! h^a_c
-\bigtriangledown^a \!\bigtriangledown^b \! h^c_c
 \right. \nn \\
&&\hspace{-3.6ex}\hspace{2ex} \left.
-\,\Box h^{ab} 
+ g^{ab} \,\bigl( \Box h^c_c \hspace{-0.1ex}-\hspace{-0.1ex}
\bigtriangledown^c \!\bigtriangledown^d \! h_{cd}
\hspace{0.2ex} \bigr) 
+ \bigl( \bigtriangledown^a h^b_c
\hspace{-0.1ex}+\hspace{-0.1ex} \bigtriangledown^b h^a_c
\hspace{-0.1ex}-\hspace{-0.1ex}
\bigtriangledown_{\!c}\hspace{0.2ex} h^{ab}\hspace{-0.1ex}
-2\hspace{0.1ex} g^{ab}\hspace{-0.2ex} 
\bigtriangledown^d \! h_{cd}
+g^{ab}\hspace{-0.2ex} \bigtriangledown_{\!c}\! h^d_d
\hspace{0.2ex} \bigr)\hspace{0.2ex} \bigtriangledown^c
\right. \nn \\
 &&\hspace{-3.6ex}\hspace{2ex} \left. 
+\,2\hspace{0.1ex} h^{ab}\, \Box
- 2\hspace{0.1ex} g^{ab}\hspace{0.2ex} 
h_{cd} \bigtriangledown^c\! \bigtriangledown^d
\hspace{0.2ex}
\right] \Phi_{n}^2.
\eea

\newpage


\end{document}